\documentclass[useAMS,usenatbib,usegraphicx]{mn2e}
\usepackage{verbatim} 
\usepackage{times}
\usepackage{amsmath,amsfonts,amssymb}
\usepackage[colorlinks=True,%
            urlcolor=blue,%
            linkcolor=blue,%
            citecolor= blue]{hyperref}

\title{Effect of shear and magnetic field on the heat-transfer efficiency of convection in rotating spherical shells}
\author[R. Yadav et al.]
  {R. K. Yadav$^{1,2}$\thanks{Email: rakesh.yadav@cfa.harvard.edu}\thanks{Current address: Harvard-Smithsonian Center for Astrophysics, 60 Garden Street, Cambridge, 02138 MA, USA}, {T. Gastine$^1$}, {U. R. Christensen$^1$}, {L. D. V. Duarte$^{3}$}, {A. Reiners$^4$} \\
  $^1$ Max-Planck-Institut f\"{u}r Sonnensystemforschung, Justus-von-Liebig-Weg 3, 37077 G\"{o}ttingen, Germany\\
  $^2$ Harvard-Smithsonian Center for Astrophysics, 60 Garden Street, Cambridge, 02138 MA, USA\\
  $^3$ Department of Physics and Astronomy, University of Exeter, Prince of Wales Road, EX4 4SB Exeter, United Kingdom \\
  $^4$ Institut f\"ur Astrophysik, Georg-August-Universit\"at, Friedrich-Hund-Platz 1, 37077 G\"ottingen, Germany
  }
\journal{Geophys. J. Int.}
\date{Received 1998 December 18; in original form 1998 November 22}
\pagerange{\pageref{firstpage}--\pageref{lastpage}}
\volume{200}
\pubyear{2015}

\begin{document}

\label{firstpage}

\maketitle

\begin{abstract}
We study rotating thermal convection in spherical shells as prototype for flow in the cores of terrestrial planets, gas planets or in stars. We base our analysis on a set of about 450 direct numerical simulations of the (magneto)hydrodynamic equations under the  Boussinesq approximation. The Ekman number ranges from $10^{-3}$ to $10^{-5}$. The supercriticality of the convection reaches about 1000 in some models. Four sets of simulations are considered: non-magnetic simulations and dynamo simulations with either free-slip or no-slip flow boundary conditions. The non-magnetic setup with free-slip boundaries generates the strongest zonal flows. Both non-magnetic simulations with no-slip flow boundary conditions and self-consistent dynamos with free-slip boundaries have drastically reduced zonal-flows. Suppression of shear leads to a substantial gain in heat-transfer efficiency, increasing by a factor of 3 in some cases. Such efficiency enhancement occurs as long as the convection is significantly influenced by rotation. At higher convective driving the heat-transfer efficiency tends towards that of the classical non-rotating Rayleigh-B\'enard system. Analysis of the latitudinal distribution of heat flow at the outer boundary reveals that the shear is most effective at suppressing heat-transfer in the equatorial regions. Simulations with convection zones of different thickness show that the zonal flows becomes less energetic in thicker shells, and, therefore, their effect on heat-transfer efficiency decreases. Furthermore,  we explore the influence of the magnetic field on the {\em non-zonal} flow components of the convection. For this we compare the heat-transfer efficiency of no-slip non-magnetic cases with that of the no-slip dynamo simulations. We find that at $E=10^{-5}$ magnetic field significantly affects the convection and a maximum gain of about 30\% (as compared to the non-magnetic case) in heat-transfer efficiency is obtained for an Elsasser number of about 3. Our analysis motivates us to speculate that convection in the polar regions in dynamos at $E=10^{-5}$ is probably in a `magnetostrophic' regime.
\end{abstract}


\section{Introduction}
The magnetic field we see on the Earth, on other planets and stars, is thought to be generated via a dynamo mechanism which is powered by the convective motions in the interior of these objects~\cite[e.g.][]{brandenburg2005, donati2009, wicht2010,  jones2011, roberts2013}. In terrestrial planets like the Earth, the convection is driven by a combination of compositional and thermal buoyancy, while for gas planets and stars convection of thermal origin is a more efficient means of heat transport than conduction or radiation.  

Due to the advances in the computing power numerical simulation of the equations that govern convective flow has become an important tool for enhancing our understanding of convection in natural objects. A lot of effort has been dedicated to studying thermal convection in the simplified setup of a plane layer where the bottom boundary is hotter than the top boundary~\cite[e.g.][]{ahlers2009}. This idealized form of convection is referred to as  {\em Rayleigh-B{\'e}nard} convection (RBC). Many studies have focused on the heat-transfer efficiency of the convection as a function of the various physical parameters of the system. An understanding of such a relationship allows us to model the thermal evolution of planets and stars~\citep[e.g.][]{breuer2010, kippenhahn2012}, and can help to elucidate the nature of the dynamo operating in such systems.

Earth, other planets, and stars rotate and the associated Coriolis forces affect the properties of the convection occurring in their interior. The Proudman-Taylor theorem~\citep{proudman1916, taylor1923} demonstrates that in rotating systems, Coriolis forces tend to suppress motion in the direction along the rotation axis, rendering the flow nearly two-dimensional. If the RBC setup described above rotates along the vertical direction, convection occurs in columnar cells aligned with the rotation axis. The vertical component of the flow, necessary for heat transfer, will be significantly reduced.

The presence of magnetic field is ubiquitous in planets and stars. It is well known that imposing a magnetic field suppresses the convective motions via the associated Lorentz forces in a typical RBC setup~\citep{chandrasekhar1961}. Magnetic quenching can reduce the heat-transport efficiency of the convecting flow. The effect of the dynamo-generated magnetic field on the convective flow is more complex than that of a uniform imposed field and depends on various factors, for example, the rotation rate of the system, the degree of density stratification in the convecting layer, etc.

The interplay of the effect of rotation and magnetic field on the convection is of particular importance as it relates to the situation in planets like the Earth which rotate rapidly and possess a strong dynamo-generated magnetic field~\citep{roberts2013}. Acting alone, both rotation and magnetic field quench the convection. However, under some conditions, the presence of magnetic field in a rotating system can enhance the heat-transport efficiency as compared to the corresponding rotating setup without magnetic field~\citep{chandrasekhar1961, stevenson1979}. This `magnetostrophic' regime, in which Lorentz force and Coriolis force are (aside from pressure forces) dominant in the force balance of the flow, is thought to be of fundamental importance for the Earth and for other rapidly rotating planets~\citep{jones2011, roberts2013, davidson2013}.

Geometry can also play a crucial role in determining the behaviour of convecting fluids. It is well known that convection in rotating  spherical shells can self-consistently excite strong axisymmetric azimuthal (zonal) flows~\citep{gilman1977}, with free-slip mechanical boundary conditions being especially suitable for permitting such flows~\citep{gilman1978}. The excitation of zonal flows in rotating spherical shells by Reynolds stresses is thought to be due to the effect of the curved spherical boundaries on the convection~\citep{busse1982}. In non-rotating two dimensional RBC with horizontally periodic boundaries similar shear flow can be excited via the so-called `tilting' instability of convection rolls~\citep{thompson1970, fitzgerald2014}. Although, the non-rotating three-dimensional (3D) RBC with periodic boundaries does not generate strong zonal flows. In such systems, even if localized tilting of convection cells excite some zonal flow, similar processes in neighbouring locations may suppresses it since the convective rolls might be randomly oriented. However, such 3D setups can promote strong zonal flows if they rotate along a horizontal direction~\citep{hardenberg2015}. Apart from not contributing to heat-transfer, such zonal flow is capable of deflecting and shearing the heat-transferring flow itself~\citep{terry2000, goluskin2014}.

The effects described above highlight that (magneto)hydrodynamic convection in rotating spherical shells is richer in physical phenomena than the classical RBC setup. Rotating RBC has been investigated using both theoretical and laboratory experiments~\cite[see e.g.][and references therein]{king2012, ecke2014, cheng2015}. One of the main results of these studies is the existence of a transition regime~\citep[see][]{aurnou2015}. Depending on the rotation rate and other control parameters of the system, the Coriolis forces can either strongly affect the flow and hence the heat-transfer efficiency of convection, or other forces (buoyancy or inertia) can dominate and the system approaches the non-rotating RBC state. The exact relationship describing the dependence of the heat-transfer efficiency on the physical parameters of the system is still a matter of debate. Rotating RBC with either an imposed magnetic field or a self-generated one has also been studied. \citet{stellmach2004} reported that  applied/self-generated magnetic field could substantially change the length scale and vigor of the convection. Recently, \citet{king2015} experimentally investigated rotating RBC with an imposed magnetic field and also reported enhancement of convective efficiency depending on the applied magnetic field.

Rotating and magnetic convection in spherical shells has been investigated in the context of geodynamo models. Here, the focus is usually on characterising the dynamo-generated magnetic field. \citet{christensen1999} investigated a set of spherical dynamo simulations and explored the effect of various control parameters on the strength and morphology of the magnetic field. There have been more extensive systematic parameters studies since then which have tried to focus on various other relevant aspects. For example, the effect of the fluid Prandtl number~\citep{simitev2005} and the effect of different velocity and temperature boundary conditions~\citep{christensen2002, busse2006, willis2007, schrinner2012, davies2013, yadav2013a, dharmaraj2014}, on convection and magnetic field have been investigated using parameter studies.

In this study we take a somewhat different approach and investigate exclusively how convective {\em flow} is affected by different velocity boundary conditions, control parameters, and the presence of dynamo-generated magnetic field. We will build our study on an extensive set of numerical simulations which reach parameter values that were so far hard to access in numerical simulations and cover a wide enough parameter space to highlight a few interesting and novel trends. The basic setup of the majority of the simulations in the data-set is similar to the typical geodynamo models studied in the past.

At this juncture it is appropriate to point out what our study does not touch upon. To determine the accurate relationship between the amount of heat transferred via convective motions and the system control parameters is of fundamental importance, but is not the aim of this study. In the context of magnetic convection in rotating spherical shells, this has been investigated by \citet{king2010}. Here we concentrate on how heat-transfer is affected by the factors discussed above: boundary conditions, zonal flow, and presence/absence of magnetic field. Furthermore, recent developments in the rotating convection theory and simulations~\citep[see e.g.][]{julien2012, cheng2015} propose control parameters for asymptotic regimes which are beyond what we can achieve with our computational resources. Nonetheless, we expect that at least the qualitative trends we find in our simulations are relevant in the asymptotic regime in planets and stars.

We describe the governing equations and the model ingredients in Sec.~\ref{model}. In Sec.~\ref{hydro_sec}, we analyse purely hydrodynamic convection in rotating spherical shells with different boundary conditions. We then discuss self-consistent dynamo simulations in Sec.~\ref{dynamo_sec} and use two different setups with free-slip boundaries and no-slip boundaries to disentangle the effect of magnetic field on zonal and non-zonal flow components. We conclude and discuss possible extensions of our study in Sec.~\ref{sum}.

\section{Model setup} \label{model}

\subsection{MHD Equations}
We solve the fundamental (magneto)hydrodynamic  (MHD) equations in a rotating spherical shell containing an incompressible fluid of constant density $\rho$. The spherical shell's  inner and outer boundaries are located at radius $r_i$ and $r_o$ respectively. The ratio $r_i/r_o$ defines the aspect ratio $\eta$ of the spherical shell. The spherical shell rotates with a frequency $\Omega$ along the vertical axis $\hat{\bf z}$. We work with a non-dimensional set of equations where the fundamental units are: the shell thickness $D=r_o-r_i$ for length scale, the viscous diffusion time $D^2/\nu$ ($\nu$ is the viscosity) for time scale , $\sqrt{\rho\mu\lambda\Omega}$ ($\mu$ is the magnetic permeability and $\lambda$ is the magnetic diffusivity) for magnetic field scale, and the imposed temperature contrast $\Delta T$ between the inner and the outer boundary for  temperature scale. 

With these non-dimensional units, the Boussinesq equations governing the velocity $\mathbf{u}$, the temperature $T$, and the magnetic field $\mathbf{B}$, are given by
\begin{gather}
E\left(\frac{\partial\mathbf{u}}{\partial t}+\mathbf{u\cdot\nabla\mathbf{u}}\right)+2\hat{z}\times\mathbf{u}= -\nabla P + \frac{Ra\,E}{P_r}\,{g(r)\,T\,\hat{\bf r}} \nonumber \\ +{\frac{1}{P_m}}(\nabla\times\mathbf{B})\times\mathbf{B}+E\nabla^{2}\mathbf{u},  \label{eq:MHD_vel} 
\end{gather}
\begin{gather}
\nabla \cdot \mathbf{u}  =  0, 
\end{gather}
\begin{gather}
\frac{\partial T}{\partial t}+\mathbf{u\cdot\nabla}T  = \frac{1}{P_r}\nabla^{2}T, 
\end{gather}
\begin{gather}
\frac{\partial\mathbf{B}}{\partial t}  =  \nabla\times(\mathbf{u}\times\mathbf{B})+\frac{1}{P_m}\nabla^{2}\mathbf{B}, \label{eq:MHD_mag}
\end{gather}
\begin{gather}
\nabla \cdot \mathbf{B}  =  0,  \label{eq:div_B_0}
\end{gather}
where $P$ is the pressure, and $g(r)$ is the gravity profile which is either $r/r_o$ or $(r_o/ r)^2$. In purely hydrodynamical simulations, Eqns.~\ref{eq:MHD_mag} and \ref{eq:div_B_0} are not relevant and $\mathbf{B}$ is set to zero in Eqn.~\ref{eq:MHD_vel}. The non-dimensional control parameters appearing in the equations above are defined as follows:
\begin{gather}
\text{Prandtl number } \,\,\,\, P_r=\frac{\nu}{\kappa},  \\
\text{magnetic Prandtl number }  \,\,\,\, P_m=\frac{\nu}{\lambda},  \\
\text{Rayleigh number }  \,\,\,\, Ra=\frac{\alpha\,g_o\,D^3\Delta T}{\nu\,\kappa}, \\ 
\text{ Ekman number }  \,\,\,\, E=\frac{\nu}{\Omega D^{2}}, 
\end{gather}
where $\kappa$ is the thermal diffusivity, $\alpha$ is the thermal expansivity, and $g_o$ is the gravity at the outer boundary. 

\subsection{Boundary Conditions}
We assume the same boundary conditions for both boundaries in the majority of the cases. This is done for the sake of simplicity and also since we are not modelling a specific planet where such a choice might not be appropriate (for instance, in the Earth, inner solid core and outer mantle impose different magnetic boundary conditions). The boundary conditions are also similar to many earlier studies (mainly in the geodynamo context) which facilitates direct comparison. 

For all of the simulations in this study the temperature is fixed on both boundaries, with the inner boundary being hotter than the outer one by $\Delta T$. For most simulations both boundaries are either free-slip or no-slip. For simulations with the deepest convection zone ($\eta=0.2$), the inner flow boundary condition is no-slip while the outer one is free-slip (data set was taken from an earlier study). The inner core and the space beyond the outer boundary are considered insulating, i.e. the magnetic field matches a potential field on both boundaries for all of the dynamo simulations. However, for cases with $\eta=0.2$ the inner core is conducting.

\subsection{Numerical Method}
We use the pseudo-spectral open source\footnote{Available at \href{https://github.com/magic-sph/magic}{\tt https://github.com/magic-sph/magic}} code MagIC \citep{wicht2002} to solve the MHD equations (\ref{eq:MHD_vel}-\ref{eq:div_B_0}) with the boundary conditions mentioned above. MagIC has been successfully tested against the community benchmarks, for both Boussinesq~\citep{christensen2001a} and anelastic~\citep{jones2011an} convective dynamo simulations. To ensure that the velocity and the magnetic field are solenoidal they are written as a combination of a poloidal and a toroidal part as
\begin{gather}
\mathbf{u}=\nabla\times\nabla\times C\hat{\bf r} + \nabla\times D\hat{\bf r},  \\
\mathbf{B}=\nabla\times\nabla\times P\hat{\bf r} + \nabla\times Q\hat{\bf r}. 
\end{gather}
The temperature, pressure, and the scalar potentials $C, D, P, Q$, are expanded in terms of the spherical harmonics for the azimuthal and the latitudinal direction and as a sum of Chebyshev polynomials for the radial direction. A combined implicit and explicit time-advancing strategy is used for time integrating the equations: an explicit second-order Adams-Bashforth scheme is used to time-advance the non-linear and the Coriolis term and an implicit Crank-Nicolson scheme is used for the other terms~\citep{glatzmaier1984, christensen2007}.

\subsection{A parameter study}
In our simulations, the Ekman number ranges from $10^{-3}$ to $10^{-5}$. The fluid Prandtl number is kept constant at unity. In most cases, the magnetic Prandtl number was also unity. Three aspect ratios, 0.2, 0.35, 0.6, are considered to exemplify very deep, moderately deep, and moderately thin convection zones. Note that the gravity is changed accordingly to better approximate the distribution of matter: $g(r)$ is $r/r_o$ in shells with $\eta$ of 0.2 and 0.35, which may represent the case of Earth's core at present and at some point in the past, while it is $(r_o/r)^2$ in shells with $\eta$ of 0.6, which could be representative of the convection zone of a solar-type star. The data for $\eta=0.35$ is available in tabulated form in the online version. More details on the cases with $\eta=0.6$ will be published in a forthcoming study.

We have tried to avoid using imposed symmetries in the azimuthal direction as much as possible, especially in the dynamo simulations. However, to expedite the initial transient stages of many demanding simulations we used either 2-fold or 4-fold symmetries in the  beginning. The results from  these simulations were then tested in full sphere calculations. The resolution requirements were checked by analysing the resulting kinetic and magnetic energy spectra (at least 3 decades of decay), by ensuring that the time-averaged total heat flux on the inner and the outer boundary match, and by matching the input buoyancy power and the dissipated power.

\section{Hydrodynamic simulations} \label{hydro_sec}
\subsection{Models with free-slip boundaries}
\begin{figure}
\includegraphics[scale=0.48]{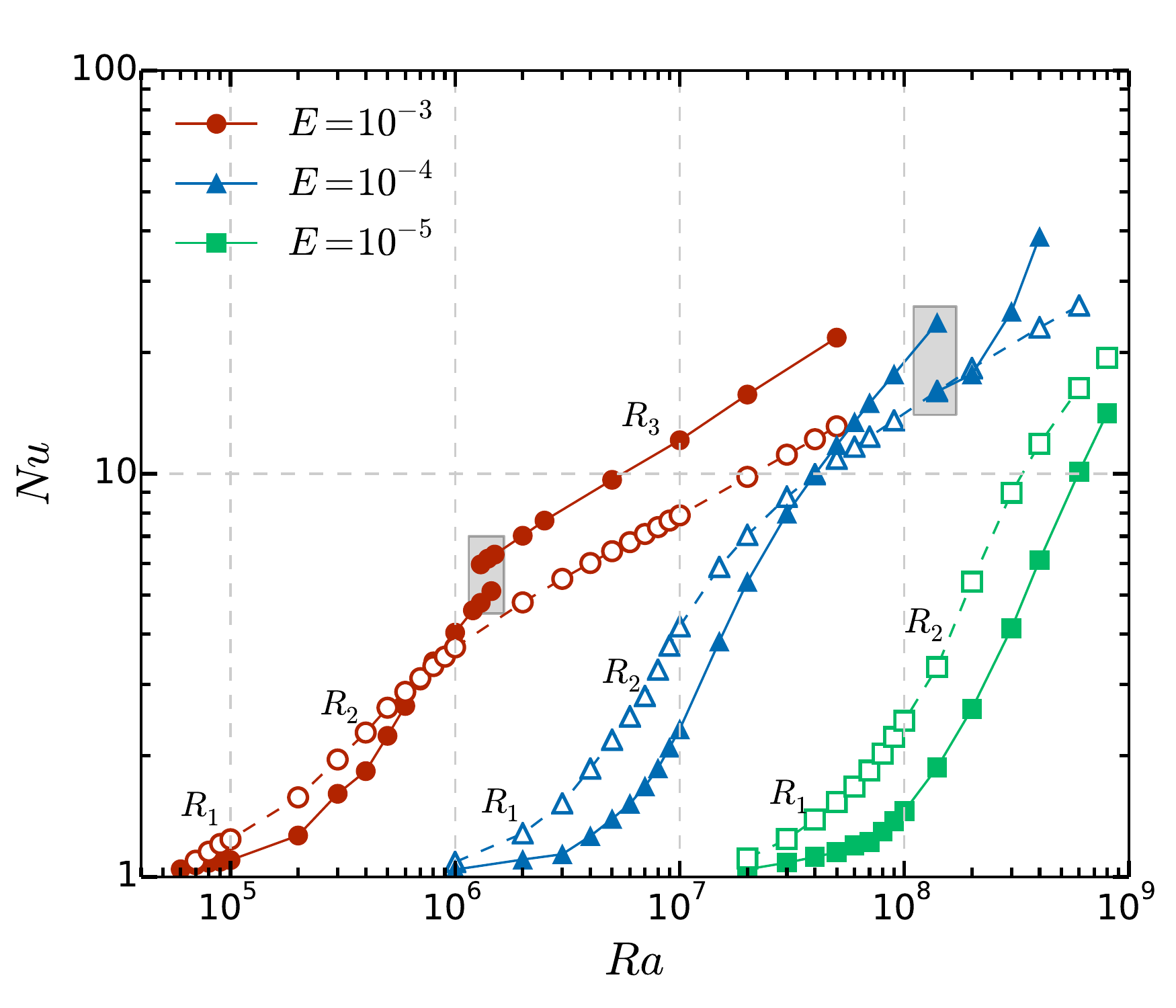}
\caption{Nusselt number $Nu$ as a function of the Rayleigh number $Ra$ for hydrodynamic simulations with free-slip boundary conditions, shown by filled data points, and hydrodynamic simulations with no-slip boundary conditions, shown by empty data points. The color/shape represents different Ekman numbers. The two gray-shaded boxes encapsulate the regime where zonal-flow bistability exists for free-slip cases. The data are for spherical shells with $\eta$ = 0.35.}
\label{Nu_FSH_NSH}
\end{figure}

The convective heat-transfer is usually measured in terms of the ratio of the total heat transported and the heat conducted when convection is absent. This quantity is called the Nusselt number whose general definition is 
\begin{gather}
Nu (r, t)=   \frac{\langle u_{r}T\rangle_{\theta,\phi}-\frac{1}{P_{r}}\frac{\partial\langle T\rangle_{\theta,\phi}}{\partial r}}{-\frac{1}{P_{r}}\frac{dT_{c}}{d r}} 
\end{gather}
where the  angular brackets denote averaging and the subscipts represent the averaging directions, and $\partial T/\partial r$ and $d T_c/d r$ are the total and conductive temperature gradients, respectively. In the following discussion we will use the time-averaged value of $Nu$ on the outer boundary of the spherical shell. Noting that the conductive temperature gradient on the outer boundary of the shell is $-\eta$, the outer boundary Nusselt number becomes 
\begin{gather}
 Nu =  -\frac{1}{\eta}\frac{\partial \langle T\rangle_{\theta,\phi}}{\partial r}\,\, (\textrm{at }r=r_o). 
\end{gather}

We begin by considering the heat-transfer behaviour of hydrodynamic (HD) convection with free-slip boundaries in a spherical shell with aspect ratio 0.35. The classical $Nu$ vs. $Ra$ behavior is shown in Fig.~\ref{Nu_FSH_NSH} for these cases with filled data points. For $E$=$10^{-3}$ and $10^{-4}$ roughly three regimes can be noticed. Just after the onset of convection (i.e. where $Nu$ starts becoming larger than 1), $Nu$ rises very slowly  as the supercriticality of the convection increases. Marker `$R_1$' highlights this regime at different Ekman numbers. Here the effect of the inertial forces are rather small. This is usually referred to as the weakly non-linear regime where a scaling of type $Nu-1 \propto (Ra-Ra_c)/Ra_c$, where $Ra_c$ is the Rayleigh number at which the convection starts, holds~\citep[e.g.][]{gillet2006, jones2007}. This regime is peculiar to spherical shells and does not exist in plane layer rotating RBC; see, for example, Figure 5 in \citet{cheng2015}. The cause for the low efficiency of heat transfer over some range of Rayleigh numbers is that near the onset of convection, columnar cells are arrayed in the vicinity of the tangent cylinder (an imaginary cylinder aligned with the rotation axis, around the inner non-convective region) and do not fill the entire shell volume~\citep{jones2000, dormy2004}. Hence conduction must transport the heat over a large part of the distance between inner and outer boundary. 

Regime $R_1$ is gradually taken over by a different scaling behaviour of the Nusselt number where it rises strongly as $Ra$ increases. This regime is marked with `$R_2$'.  The extent of $R_2$ is the largest in $E$=$10^{-5}$ data. A best-fit power law of form $Nu\propto Ra^{\alpha}$ gives $\alpha\approx 1.23$ for data with $Ra\ge 2\times10^8$ and $E$=$10^{-5}$. This is close to the scaling law $Nu\propto Ra^{6/5}$ proposed by \cite{king2010} for heat-transfer in rotating spherical-shell convection. As we increase  $Ra$ further another transition to a different regime occurs where the scaling behaviour of the Nusselt number becomes less steep, indicated  with marker `$R_3$'. In this regime, the inertial forces start to dominate over the Coriolis forces. For data with $Ra\ge 2\times10^6$ and $E$=$10^{-3}$, the power law exponent $\alpha$ is about 0.35, steeper than the value $2/7$~\citep[e.g.][]{king2010} and $1/3$~\cite[e.g.][]{ahlers2009}. This suggests that our $R_3$ regime for  $E$=$10^{-3}$ is not asymptotic but rather a `bridging' one; for higher $Ra$, the slope should change to values expected in non-rotating RBC, i.e. $1/3$, or even to $1/2$ for very high $Ra$~\citep{ahlers2009, gastine2015}.

\begin{figure*}
{\Large Non-magnetic cases with free-slip boundaries}\\
\includegraphics[scale=0.3]{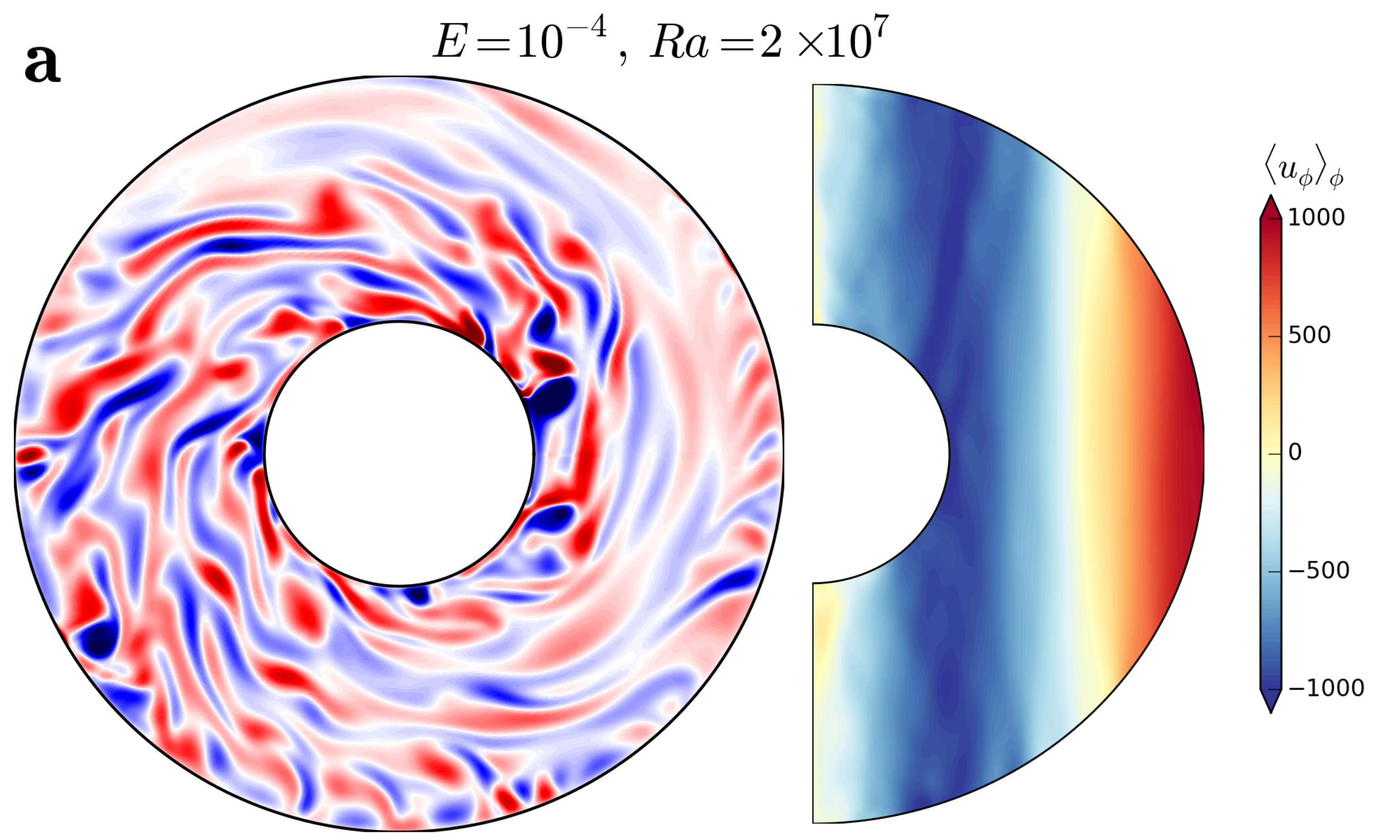} \hspace{7mm} \includegraphics[scale=0.3]{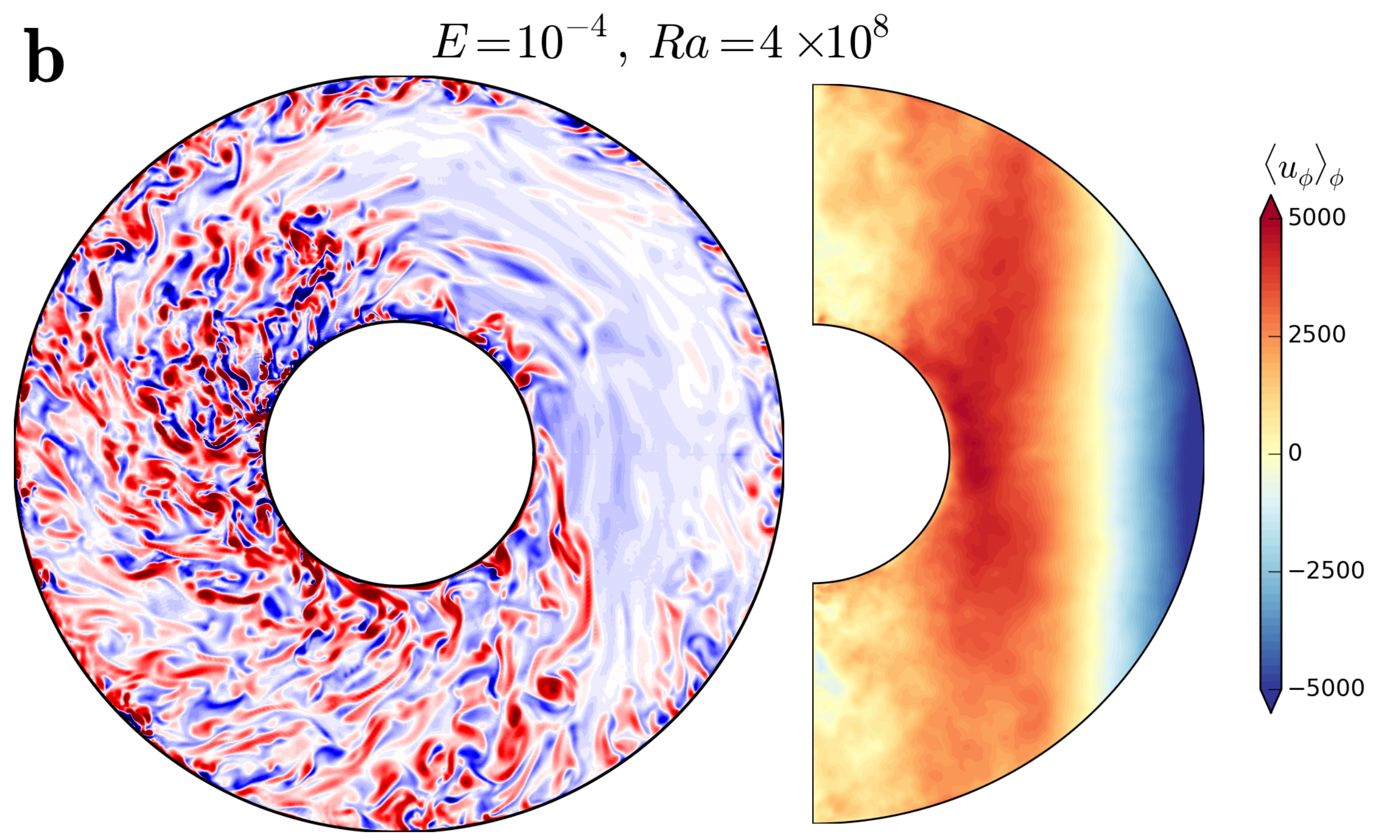}  \\
\vspace{5mm}
{\Large Non-magnetic cases with no-slip boundaries}\\
\includegraphics[scale=0.3]{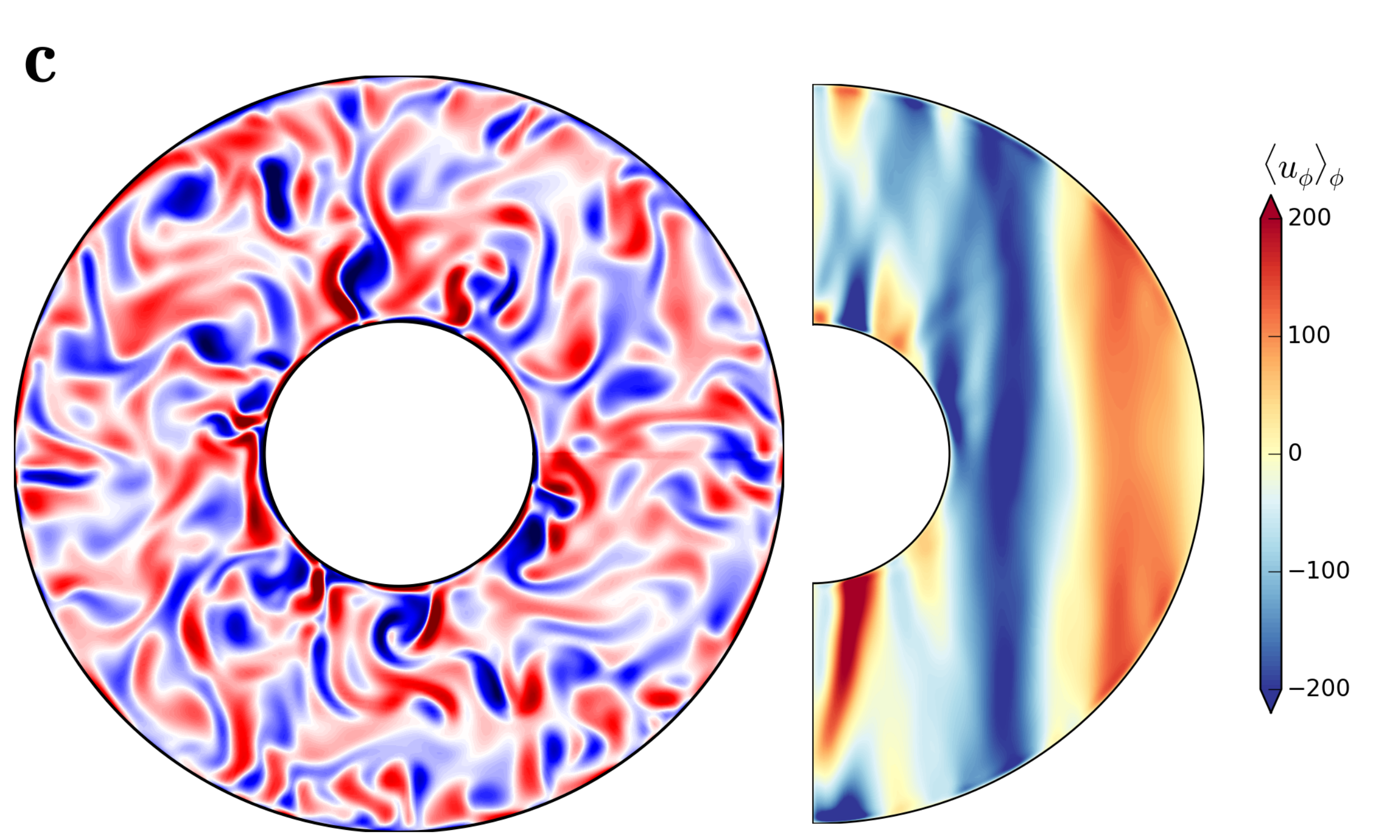} \hspace{7mm} \includegraphics[scale=0.3]{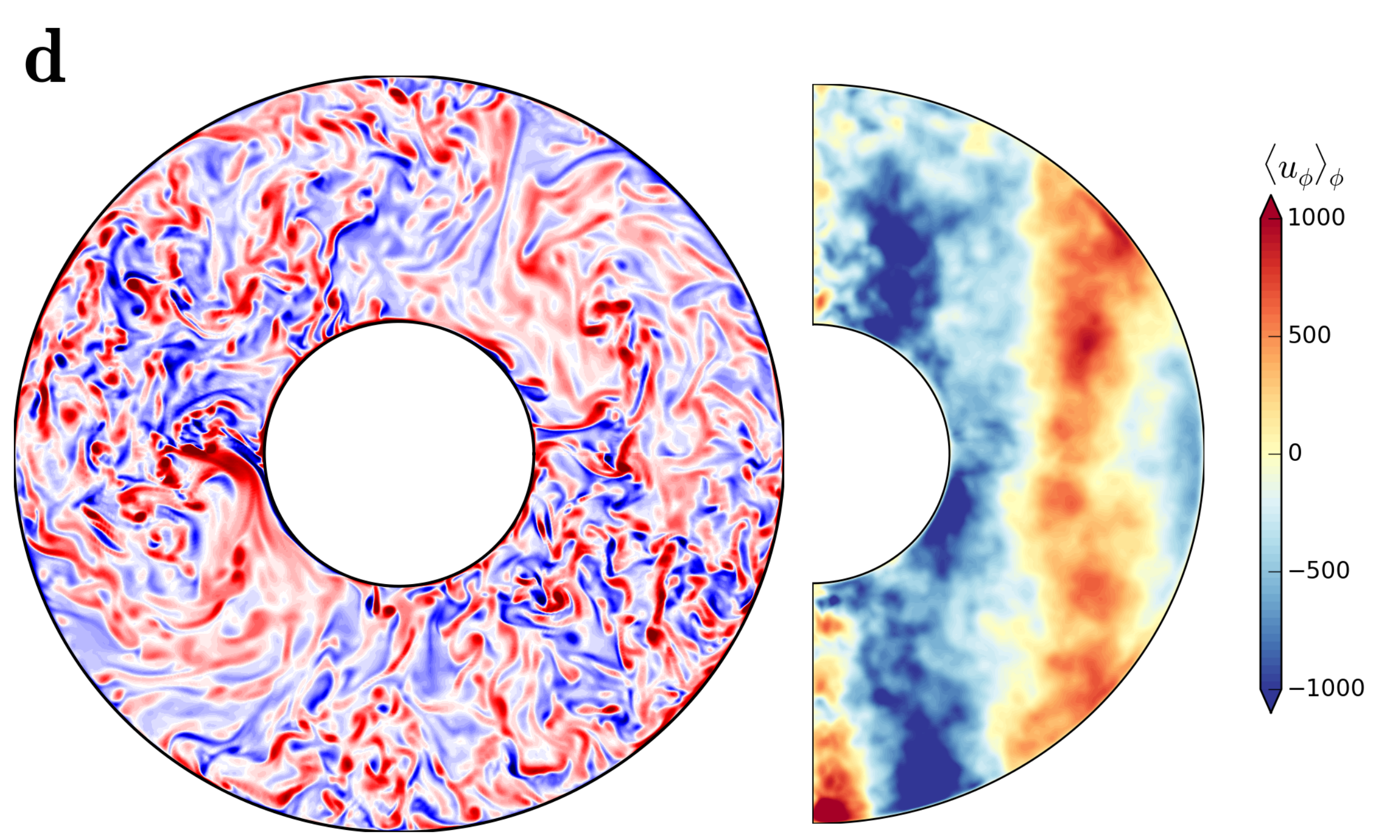}  \\
\vspace{5mm}
{\Large Dynamo cases with free-slip boundaries}\\
\includegraphics[scale=0.3]{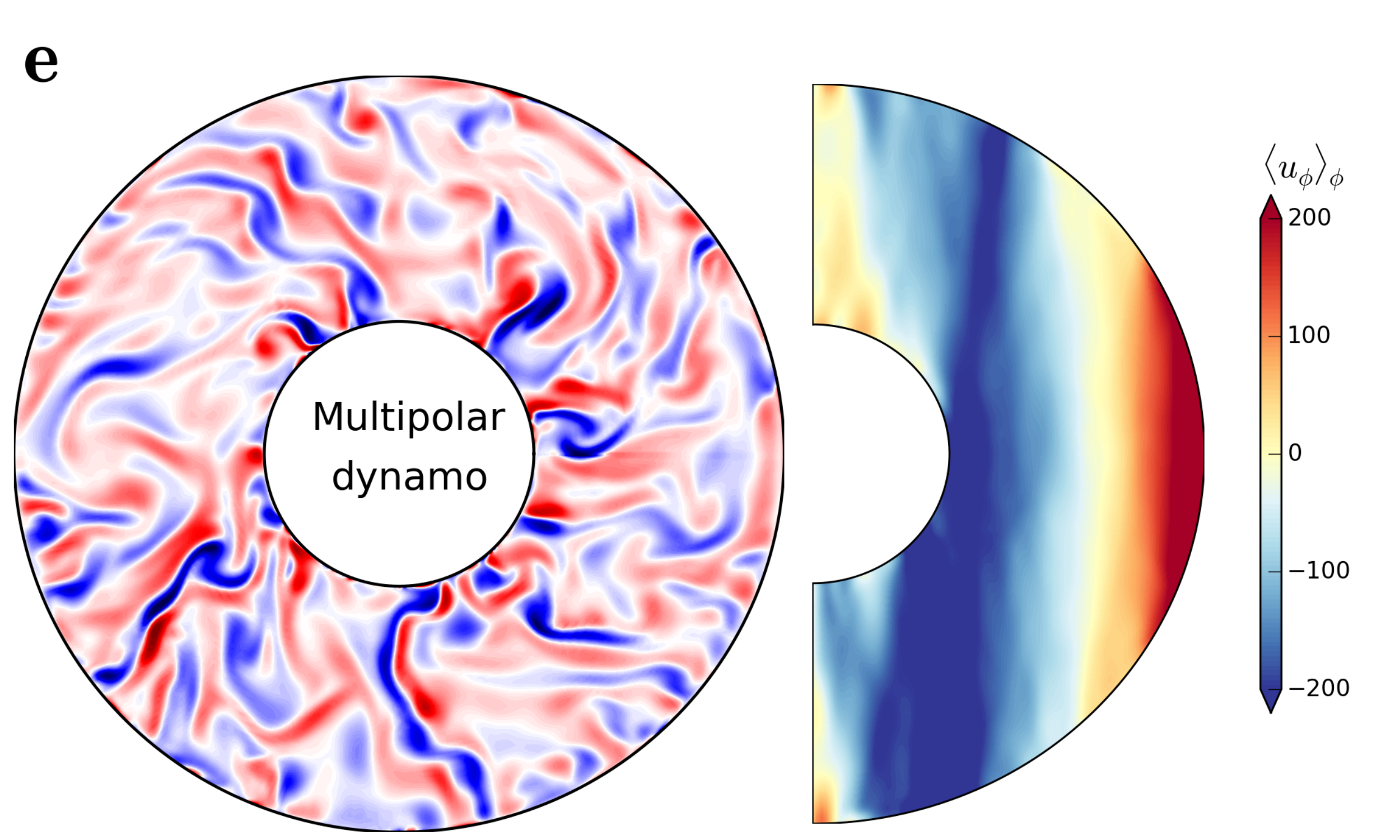} \hspace{7mm} \includegraphics[scale=0.3]{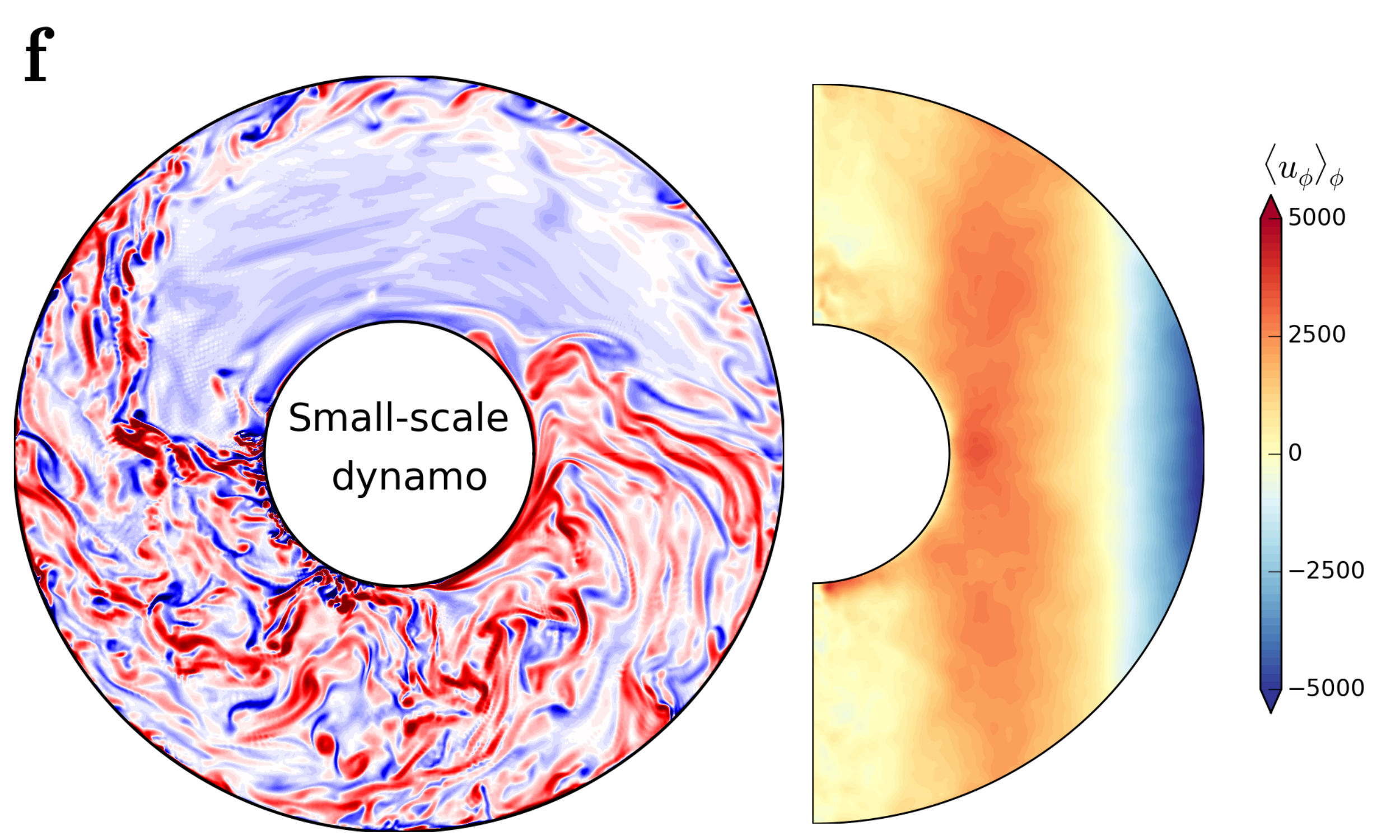} \\
\vspace{5mm}
{\Large Dynamo cases with no-slip boundaries}\\
\includegraphics[scale=0.3]{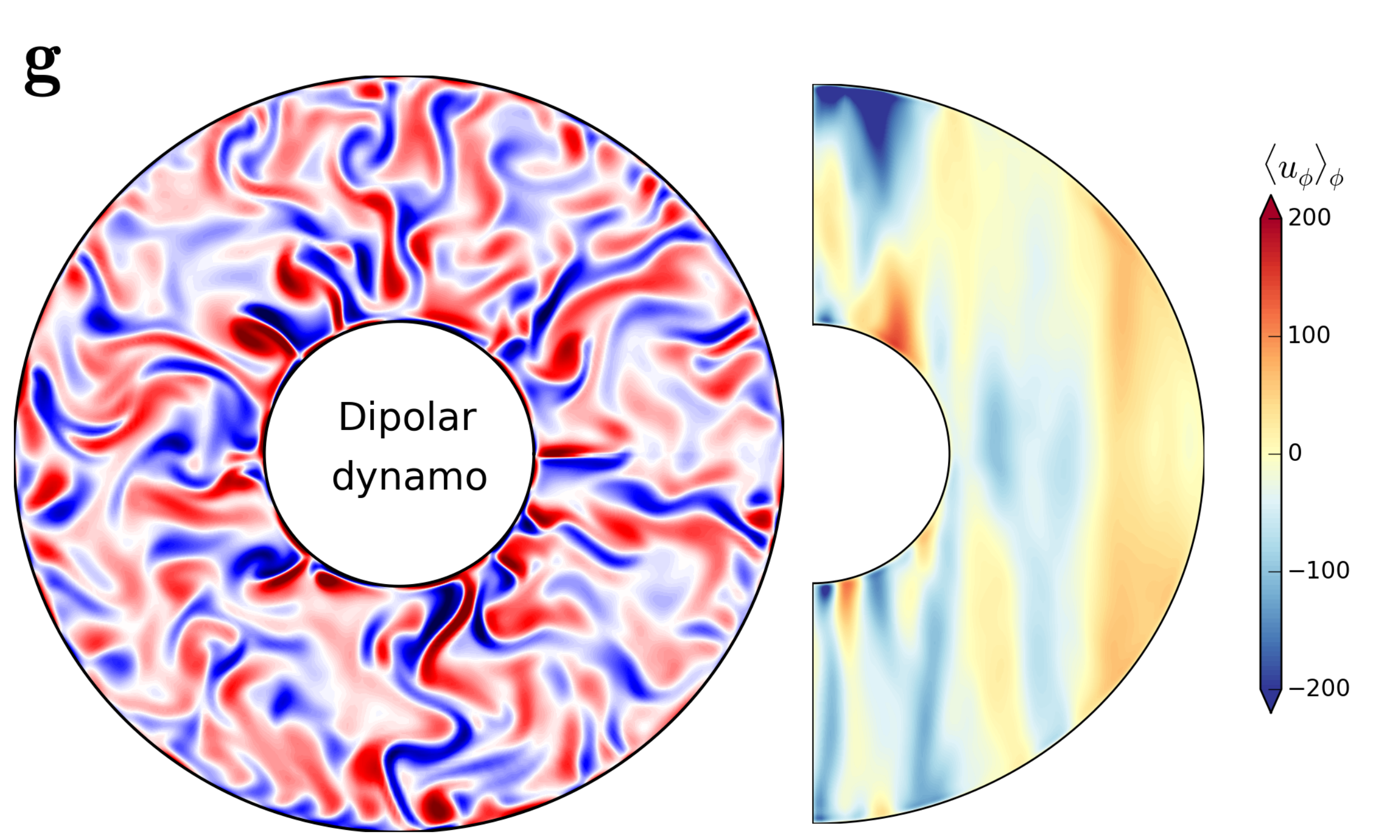} \hspace{7mm} \includegraphics[scale=0.3]{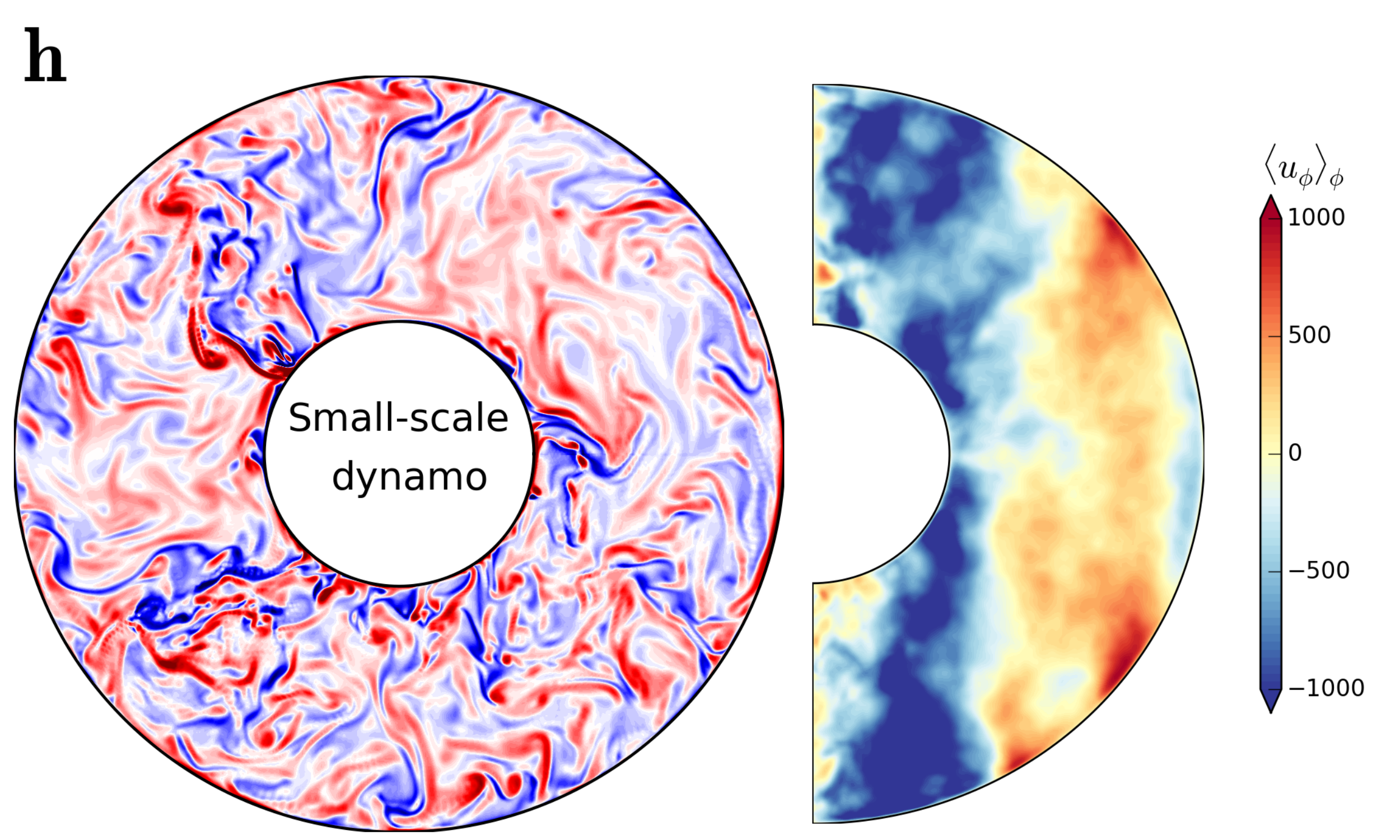} 
\caption{Snapshots of axial vorticity in the equatorial plane and azimuthally averaged zonal flow in a meridional plane.  The  zonal flow velocity is expressed in terms of a  Reynolds number. The shell is rotating anti-clockwise in the vorticity panels. The red and blue vorticity shades represent positive and negative values, respectively. In all the plots the color scale saturates below the extreme values to highlight fainter structures. Assuming a critical $Ra$ of $7\times10^{5}$ \citep{christensen2006} the supercriticality of convection is about 30 and 600 in the left and right column, respectively. \label{flow_struct}}
\end{figure*}

\subsubsection{Flow structure}
In Fig.~\ref{flow_struct}(a, b) we display the axial vorticity $(\nabla\times\mathbf{u'})_z$ in the equatorial plane (as seen from a northern viewpoint) and azimuthally averaged zonal flow $\langle u_{\phi}\rangle_{\phi}$ in a meridional plane for two exemplary non-magnetic cases with free-slip mechanical boundaries.  Note that we have used the non-zonal velocity $\mathbf{u'} = \mathbf{u}-\hat{\phi} \langle u_{\phi}\rangle_{\phi}$; this is done to isolate the vorticity contribution of the convective flow component. In rotating spherical shells, where convection is significantly influenced by the Coriolis forces, the convection takes the form of alternating cyclonic and anti-cyclonic vortical columns~\citep[for a review, see][]{jones2007, aurnou2015}. Equatorial manifestations of such vortical columns are clearly visible in the axial vorticity plot in Fig.~\ref{flow_struct}a which shows adjacent regions of positive and negative vorticity. In this regime, the columns roughly maintain their identity in $\hat{z}$-direction until they hit the outer boundary. The prograde tilt (i.e. clockwise spiralling morphology) of these columns is also apparent in the vorticity plot. 

In Fig.~\ref{flow_struct}a, the  zonal flow profile is almost invariant along the rotation axis and the flow near the equator is rotating eastward (prograde). Such kind of  zonal winds have been investigated extensively in the past, both analytically~\citep[e.g.][]{zhang1992, busse1994} and numerically~\citep{gilman1977, miesch2000, grote2000a, aurnou2001, christensen2001b, christensen2002, heimpel2005, ballot2007, kapyla2011b, gastine2012a, guerrero2013, hotta2015a}. The gas planets Jupiter and Saturn~\citep{porco2003, porco2005}, and the Sun~\citep{thompson2003} belong to this category. Significant prograde zonal flows might also be present in the Earth's outer core near the equator~\citep{gillet2015}. These zonal winds are excited due to the tilted columns of convection (seen in the vorticity plot) which arise due to the combined effect of Coriolis forces and the curvature of the outer boundaries in a spherical shell~\citep{busse1983, zhang1992, busse1994}; however, also refer to \cite{takehiro2008} for a different interpretation. Such a convection pattern introduces the so-called Reynolds stresses where a statistical correlation between the azimuthal and the cylindrically-radial component of the flow gives rise to a transport of angular momentum towards the equatorial regions close to the outer boundaries. In passing we note that for rapidly rotating convection with high supercriticality (approaching a `quasigeostrophic' state) the Rhines scale determines the number of zones and amplitude of the zonal flow~\citep{rhines1975, gastine2014b}. Here too, the curvature of the spherical shell enters via the so called ``topographic $\beta$ effect".

There are other mechanisms through which strong zonal flows can be excited. Systems with significant latitudinal variation of temperature/heat-flux can generate strong zonal flows via the `thermal-wind' mechanism~\citep[e.g.][]{sreenivasan2006a, miesch2006}. The so-called `compressional torque' effect might also be an important ingredient in generating strong zonal flows in compressible convection~\citep{glatzmaier1981, glatzmaier1982, evonuk2008, glatzmaier2009, verhoeven2014}; although, recent 3D anelastic simulations in spherical shells question the importance of this effect~\citep{gastine2014b}.

The vorticity structures are tilted in the opposite direction in panels Fig.~\ref{flow_struct}b, where the Rayleigh number is strongly enhanced compared to the simulation shown in Fig.~\ref{flow_struct}a. Moreover, the coherent structures with alternating cyclonic/anti-cyclonic vorticity seen in Fig.~\ref{flow_struct}a are no-longer present. A rather large region of predominantly anticyclonic vorticity (in blue) is visible in Fig.~\ref{flow_struct}b whereas more fractured and chaotic structures fill other regions of the shell. The zonal flow in Fig.~\ref{flow_struct}b has changed sign and is now westward (retrograde) in low-latitude regions near the outer boundary. In this regime of vigorous convection  the angular momentum is homogenized throughout the spherical shell~\citep{aurnou2007, gastine2013b}. The ice giant planets Uranus and Neptune might beloing to this category as they both show retrograde zonal flows~\citep[e.g.][]{fry2012, martin2012}. At this stage the Coriolis forces have a weaker influence on convection and the inertial effects become dominant. For even higher $Ra$, the convection eventually becomes isotropic without any appreciable axisymmetric flow. This was indeed the case of the largest $Ra$ cases with $E$=$10^{-3}$.

\subsubsection{Strong shear}
As discussed above zonal flows are a rather prominent feature in rotating convection in spherical shells. The competing influence of the buoyancy forces and the Coriolis forces dictate what would be the nature of the zonal-winds. \citet{gilman1977} introduced the so called convective Rossby number $Ro_{c} = \sqrt{Ra\,E^{2}/Pr}$ which roughly quantifies the ratio  of these two competing forces. It is instructive to analyze the energy contained in the axisymmetric zonal flows in the setup discussed above. As a measure of the relative contribution of the zonal flows we use
\begin{gather}
\varpi = \frac{1}{1-\frac{{E}_{kin}^{Z}}{E_{kin}^{tot}}} = \frac{E^{tot}_{kin}}{E^{NZ}_{kin}}, \label{zonal_ratio}
\end{gather}
where $E^{Z}_{kin}$, $E^{NZ}_{kin}$, and $E^{tot}_{kin}$ is the kinetic energy contained in the axisymmetric zonal flow component, in the non-zonal flow component, and in the total flow, respectively~\citep{christensen2002, gastine2012a}. 

\begin{figure}
\includegraphics[scale=0.48]{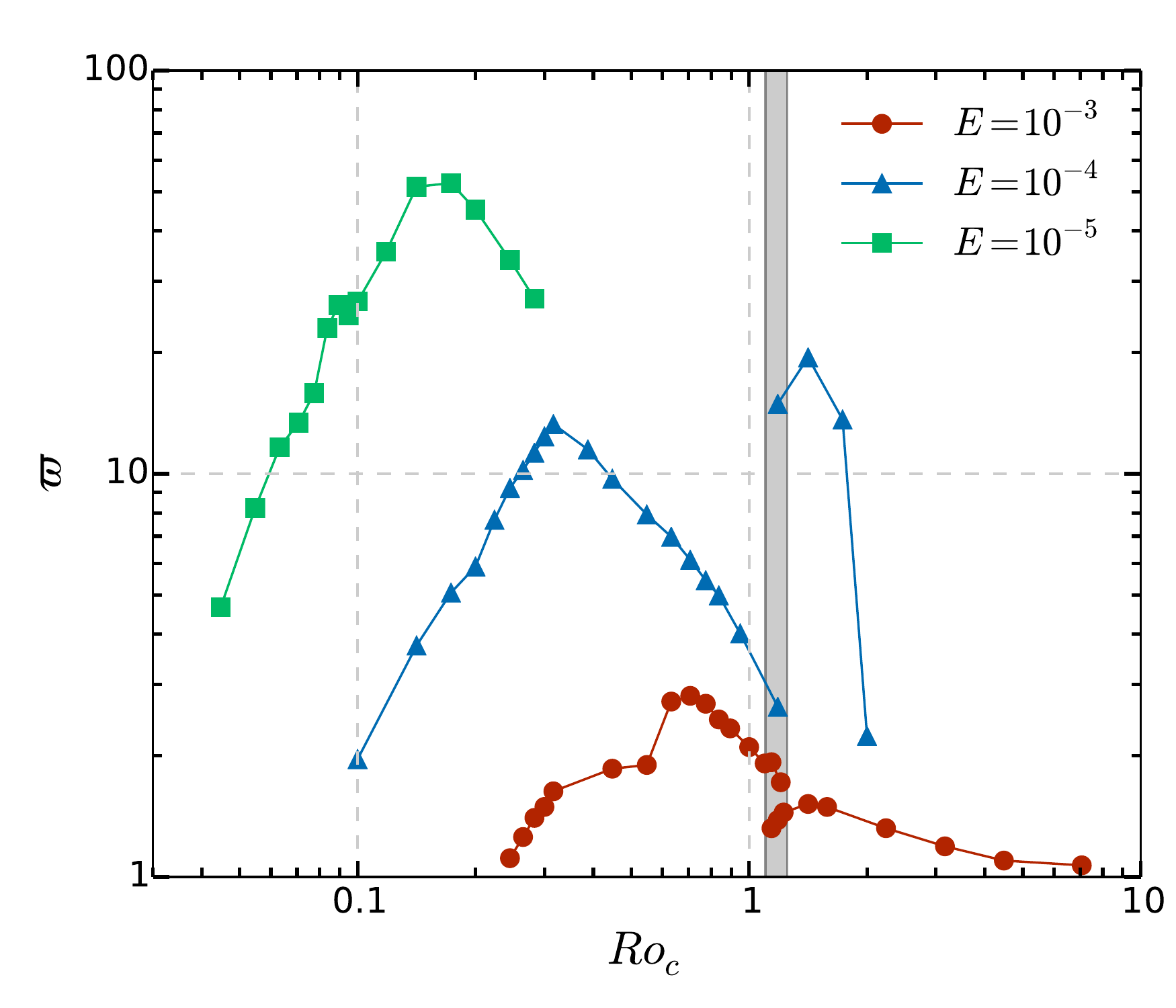}
\caption{Variation of $\varpi$ as a function of the convective Rossby number $Ro_c$ for hydrodynamic cases with free-slip boundaries. The shaded zone highlights the zonal flow bistability region.  \label{zonal_KE_FSH}}
\end{figure}

Variation of $\varpi$ as a function of $Ro_c$ is portrayed in Fig.~\ref{zonal_KE_FSH} for the non-magnetic cases with free-slip boundaries. As $Ro_c$ increases for each Ekman number, an increase in $\varpi$ shows that fraction of energy in the zonal flow increases after the onset of the convection. After reaching a peak value $\varpi$ starts declining. This may be attributed to the gradual disruption of the coherent convective columns when $Ro_c$ increases, leading to reduced Reynolds stresses which transfer energy from small-scale convective motions to large-scale azimuthal flow~\citep{christensen2002, gastine2012a}. Around $Ro_c \simeq 1$ a discontinuous change occurs in $\varpi$ for $E$=$10^{-3}$ and $10^{-4}$, after which it starts to decline again. These trends in zonal flow strength have been observed in earlier simulations as well~\citep{christensen2002, aurnou2007, gastine2012a, gastine2013b, gastine2014}.  The sudden change in $\varpi$ observed around $Ro_c\approx 1$ is due to the reversal of the prograde jet at lower $Ro_c$ to a retrograde one at higher $Ro_c$. The zonal flow energy in the retrograde-jet case exceeds that of the maximum energy in the prograde-jet case for  $E$=$10^{-4}$, consistent with \citet{gastine2013b}. However, the cases with $E$=$10^{-3}$ shows the opposite trend. It is likely due to the large Ekman number in this setup, implying a more dominant role of viscosity.

Convection in two dimensions with free-slip boundaries and one periodic direction is known for generating highly energetic `winds' or shear-flows in the direction of periodicity. Such winds are highly efficient at reducing the transport of heat perpendicular to its own direction~\citep[see e.g.][]{terry2000, garcia2003, goluskin2014}. These systems are very much similar to our setup: the tendency of two dimensionality is introduced by the influence of the Coriolis forces on the convection, the boundaries are similarly free-slip, and the azimuthal direction in spherical shell is akin to the periodic boundaries in Cartesian boxes. Therefore, it can be expected that shear might have a  strong influence on the heat transfer in our simulations as well. The axisymmetric zonal-jets do not participate in radial heat transfer, and the more energy they contain the better they become at disrupting heat-transporting motions which might try to cross them.

\subsubsection{Zonal flow bistability}
With the qualitative idea of stronger zonal flows leading to less efficient heat transfer in mind, we turn back to a feature visible in Fig.~\ref{Nu_FSH_NSH} which has not been discussed so far. For Ekman number of $10^{-3}$ and $10^{-4}$, regions where two different solutions coexist for cases with free-slip boundaries, are highlighted with gray-shaded boxes in Fig.~\ref{Nu_FSH_NSH}. In a recent study dedicated to stellar differential rotation  \cite{gastine2014} demonstrated the possibility of bistability in the nature of the differential rotation in rotating spherical shells with aspect ratio of 0.6. This result obtained in Boussinesq convection was later confirmed by fully compressible simulations~\citep{kapyla2014}; however, see \citet{karak2015}. We investigate this feature for our simulation setup as well. Prograde zonal jet persists over a wide range of low and intermediate values of the Rayleigh number, but at a sufficiently high-enough $Ra$, the equatorial zonal-jet reverses direction and becomes retrograde. If we now take this state with retrograde zonal-jet and use it as an initial condition for a simulation with a  smaller $Ra$ the retrograde zonal-jet still persists and appears to be a stable feature. The reversal of the zonal-jet and the bistability phenomenon occurs when $Ro_c\approx 1$, as highlighted in Fig.~\ref{zonal_KE_FSH} with a shaded region. In analogy with typical non-linear systems with hysteresis, there are two possible attractors for a given set of parameters in the bistable regime, and depending on the initial conditions a simulation can either be captured by the prograde attractor or the retrograde one. This phenomenon only occurs in a rather limited $Ra$ range. 

For $E$=$10^{-3}$, the retrograde-jet solution has somewhat smaller energy in the zonal flow, while, for  $E$=$10^{-4}$, the prograde-jet solutions carries less energy in the zonal flow. Therefore, it appears that more energy in the retrograde-jet solution for $E$=$10^{-4}$ leads to a lower Nusselt number as compared to the corresponding prograde-jet solutions. This scenario is reversed for $E$=$10^{-3}$ simulations. Therefore, the difference in zonal flow energy in Fig.~\ref{zonal_KE_FSH} qualitatively explains the different $Nu$ in the gray-shaded boxes in Fig.~\ref{Nu_FSH_NSH}.

\subsubsection{Latitude dependent Nusselt number}
\begin{figure*}
\includegraphics[scale=0.47]{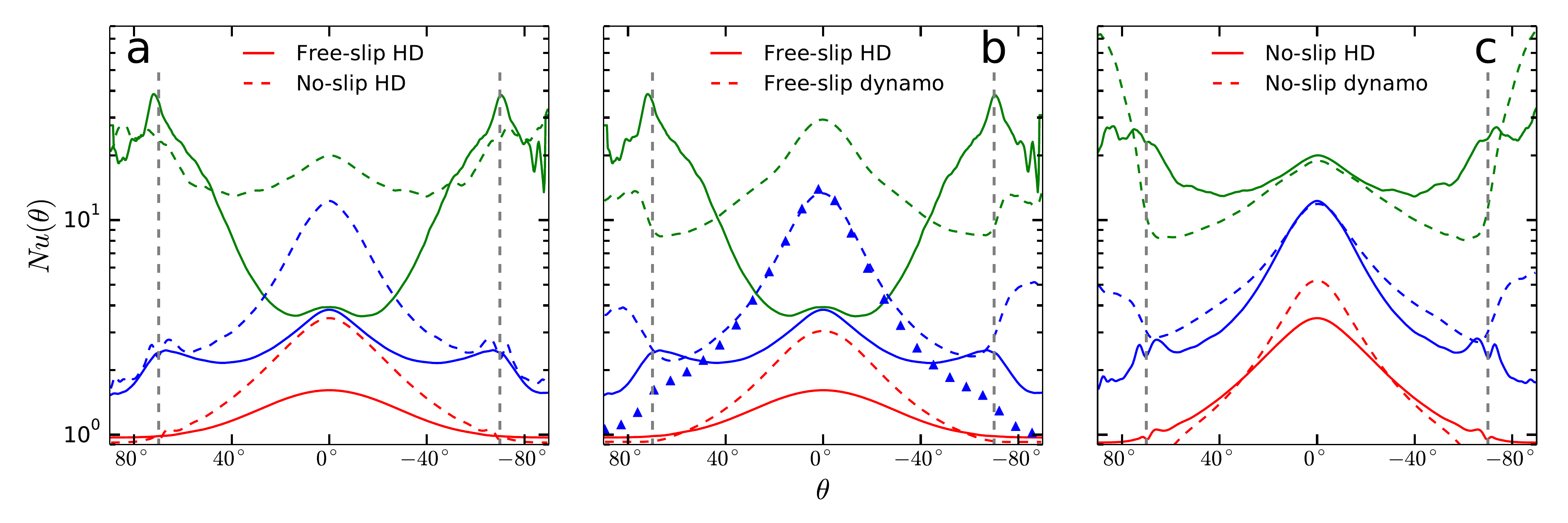}
\caption{Variation of the latitude-dependent Nusselt number $Nu(\theta)$ (see Eq.~\ref{nu_theta}) at the outer boundary for simulations with Ekman number of $10^{-5}$. Different colors in each panel represent different Rayleigh numbers: red for $Ra$=$8\times10^{7}$, blue for $Ra$=$2\times10^{8}$, and green for $Ra$=$6\times10^{8}$. The trail of blue triangular points represent the bistable multipolar dynamo solution for the same control parameter as those used for the blue dashed curve (dipolar solution). The vertical broken gray-lines represent the latitudes where the tangent cylinder touches the outer boundary. Assuming a critical $Ra$ of $10^{7}$ \citep{christensen2006} the supercriticality of convection is about 8, 20, 60 for the three $Ra$ mentioned above. \label{lat_HF}}
\end{figure*}

The distribution of heat flow with latitude on the outer boundary provides interesting information on the flow regimes inside and outside the tangent cylinder (TC). In Fig.~\ref{lat_HF}a we plot a latitude-dependent Nusselt number 
\begin{gather}
Nu(\theta) =  -\frac{1}{2\pi\eta}\int^{2\pi}_0 \frac{\partial T}{\partial r}\,d\phi \,\,\,\,\,\,\,(\textrm{at }r=r_o) \label{nu_theta}
\end{gather}
on the outer boundary for some selected HD cases with $E$=$10^{-5}$ and free-slip boundaries (solid curves). At low $Ra$ a peak in $Nu(\theta)$ in the equatorial region (small latitudes) shows that the regions inside the TC remain rather quiescent for small convective driving~\citep{jones2000, dormy2004}. As $Ra$ increases, the heat-flux at high latitudes increases rapidly once convection inside the TC sets in. Remarkably, the heat-flux in the equatorial regions nearly saturates (compare the blue and the green solid curves in Fig.~\ref{lat_HF}a). This is due to the severe quenching of radially outward flows by the strong zonal jets. In the equatorial regions, the zonal winds are nearly perpendicular to the direction of gravity, and, hence, the maximum impact is felt in this region. As we go to higher latitudes this effect decreases. Hence, we can conclude that heat-transfer efficiency near the equator is highly affected by zonal-jets.

Figure \ref{lat_HF}a highlights another interesting aspect. The rise in heat transfer with increasing Rayleigh number proceeds much faster in the polar regions (once TC convection has started) than it does in the equatorial regions. Even though convection in the equatorial regions starts at lower $Ra$, the polar Nusselt number quickly rises above the equatorial one at higher values of $Ra$. It qualitatively implies that in a relation of type $Nu\propto Ra^{\alpha}$ (dependence on other parameters is suppressed for the sake of simplicity) the exponent $\alpha$ will be higher in the polar regions as compared to the equatorial one. It may provide a plausible explanation why $\alpha$ reported in rotating box simulations, which are suited for the polar regions in a spherical shell, is generally higher~\citep{king2012, ecke2014, cheng2015} than the one reported in a similar study in spherical shells~\citep{king2010}.

\subsection{Models with no-slip boundaries} \label{NSH}
Flow boundary conditions have a strong effect on the zonal flow strength and nature~\citep{gilman1978, christensen1999, christensen2002, aubert2005, busse2006, schrinner2012,  dharmaraj2014, garcia2014}. Here we investigate the effect of changing the mechanical boundary conditions from free-slip to no-slip on the nature of heat-transport. In Fig.~\ref{Nu_FSH_NSH} we also plot $Nu$ vs. $Ra$ for non-magnetic simulations with no-slip boundaries (empty data points). Qualitatively both no-slip and free-slip data sets follow similar trends, in that there are 3 different regimes (onset, rotation dominated, and inertia dominated). There are two important differences which can be easily identified:  in the onset and in the rotation dominated regime, the Nusselt number is higher in the simulations with no-slip boundaries, while it gradually becomes lower than the free-slip cases as $Ra$ is further increased. Differences in $Nu$ between cases with  free-slip or no-slip boundaries has been reported in earlier weakly-rotating RBC studies. In these setups a higher value of $Nu$ in the free-slip cases may be attributed to the absence of viscous boundary layers~\citep{julien1996}.

\subsubsection{Flow structure}
The axial vorticity and the zonal flow plots for the no-slip HD cases are shown in Fig.~\ref{flow_struct}(c,d).  A substantial change in both panels, as compared to their free-slip counterparts in Fig.~\ref{flow_struct}(a,b), is immediately noticeable. The much weaker zonal flows show more variation along the rotation axis in the no-slip cases. Due to the reduced zonal shear, the spiralling behavior of the vorticity structures is greatly reduced. Another point to notice is that the dichotomy of prograde  and retrograde zonal jets on the outer boundary and near the equator is much less apparent here.  The reason for the quenched zonal flows can be attributed to the presence of Ekman boundary layers with no-slip boundaries~\citep[e.g.][]{jones2007}. In simulations with free-slip boundaries the saturation of the zonal flows is controlled by the weak bulk-friction due to viscosity, while, in the case of no-slip boundaries, the much larger friction in the Ekman boundary layers ($\propto E^{1/2}$, see \cite{gillet2006, jones2007}) inhibits vigorous zonal flows more efficiently~\citep{christensen2002, aubert2005, jones2007}. 

\subsubsection{Latitude dependent Nusselt number}
Figure \ref{lat_HF}a also displays $Nu(\theta)$ for no-slip HD simulations with dashed curves. Inside the TC no-slip HD cases resemble their free-slip counterparts. However, in the low-latitude regions outside the TC, the difference is rather substantial. As discussed before, the zonal flow is quenched by the no-slip boundaries, and, therefore, it does not impede the heat transfer as efficiently as in the free-slip case. As a result, the low-latitude regions in no-slip simulations remain efficient at transporting heat.

\subsection{Relation between shear flow and heat transport}
\begin{figure}
\includegraphics[scale=0.48]{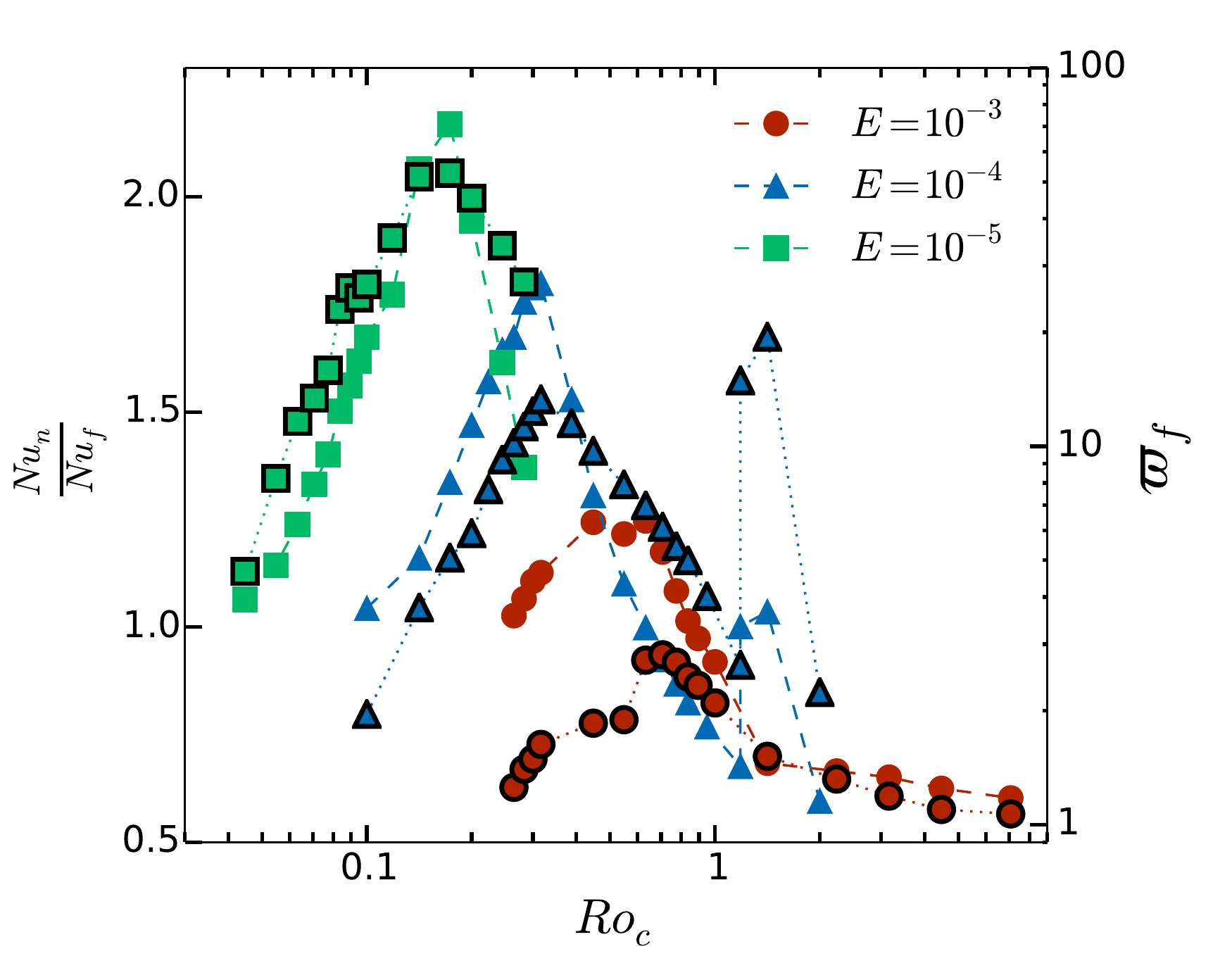}
\caption{On the left axis, ratio of the Nusselt number $Nu$ for no-slip (subscript `n') and free-slip (subscript `f') cases, shown by data points without rim, and, on the right axis, $\varpi_{f}$ (black rimmed data points) for free-slip cases as a function of the convective Rossby number $Ro_c$. The data set is for non-magnetic simulations with aspect ratio of 0.35. \label{Nu_ratio_FSH_NSH}}
\end{figure}

To show the effect of zonal flow on the heat-transfer we compare in Fig.~\ref{Nu_ratio_FSH_NSH} two relevant quantities as a function of $Ro_c$. One is the ratio of the Nusselt number in the no-slip case, $Nu_n$, to the corresponding value in the free-slip case, $Nu_f$, at otherwise the same parameter values. The other, distinguished by black-rimmed symbols, is the parameter $\varpi_{f}$ for free-slip cases (our measure for the relative contribution of zonal flow to the kinetic energy). Note that $\varpi$ for the no-slip cases remains close to unity due to quenched zonal flows. As can be clearly seen in Fig.~\ref{Nu_ratio_FSH_NSH} there is a tight correlation between the two quantities. Qualitatively, it implies that the larger the amount of energy contained in the zonal jets in the free-slip cases, the larger will be the impact on heat-transfer if we quench these jets by replacing the free-slip boundary condition by no-slip. A sudden increase in the $Nu$-ratio near $Ro_c\approx1.5$ is induced by the increase in the energy content of the zonal flow due to the transition to the retrograde-jet regime in the free-slip cases. In the large $Ro_c$ regime $\varpi$ is expected to approach unity since the system will resemble an RBC type state where any axisymmetric flow will carry negligible energy as compared to the total kinetic energy. This is starting to happen for the $E$=$10^{-3}$ cases. The $Nu$-ratio, however, will not be unity as long as the influence of the boundary layers is felt by convection; it should be smaller than unity for high $Ro_c$ since free-slip cases transport more heat~\citep{julien1996}. 

It should be noted that we have not analyzed the contribution of the Ekman pumping mechanism, which is present in no-slip cases, on the heat-transfer efficiency gain~\citep[e.g.][]{stellmach2014}. However, noting that $Nu$ was very much similar in both free-slip and no-slip HD simulations inside the TC we may speculate that this effect is smaller at our control parameters.

To summarize, in the rotation dominated regime ($Ro_c<1$), the simulations with no-slip boundaries transfer more heat because of their much weaker zonal flows. In the inertia/buoyancy dominated regime ($Ro_c>1$), however, the cases with free-slip boudnaries transfer more heat due to the lack of viscous boundary layers. 

\subsection{\label{sec:thin}Effect of shell thickness}
\begin{figure}
\includegraphics[scale=0.48]{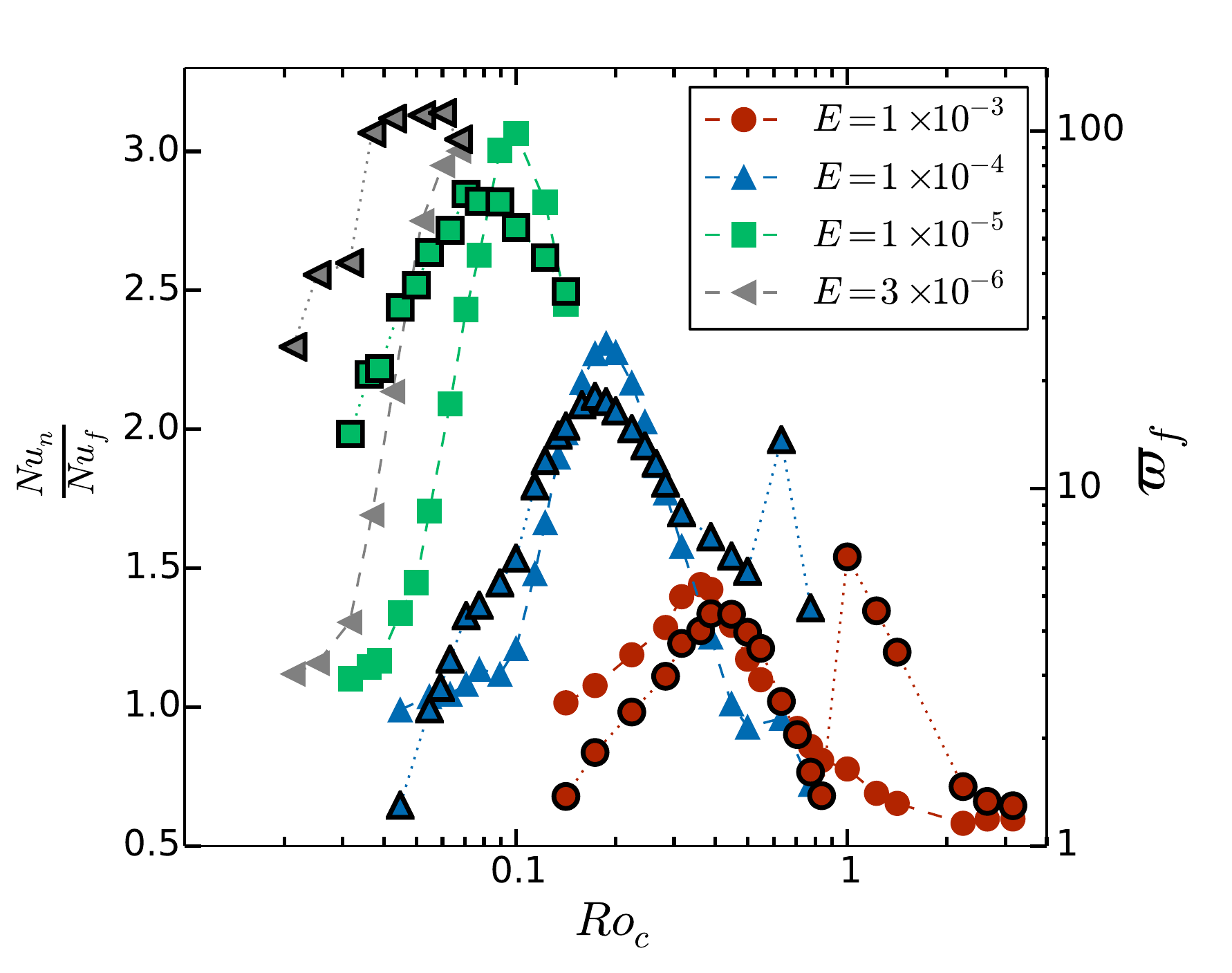}
\caption{On the left axis, ratio of the Nusselt number $Nu$ for no-slip (subscript `n') and free-slip (subscript `f') cases, and, on the right axis, $\varpi_{f}$ (black rimmed data points) for free-slip cases as a function of the convective Rossby number $Ro_c$. The data set is for non-magnetic simulations with aspect ratio of 0.6. Note the additional data points at $E$=$3\times 10^{-6}$ in gray.  \label{Nu_ratio_TFSH_TNSH}}
\end{figure}

The dominance of zonal flows appears to be somewhat enhanced in spherical shells with larger aspect ratios and free-slip boundaries~\citep[e.g.][]{aurnou2008, gastine2013b}. This would imply that the difference in the Nusselt number between free-slip and no-slip simulations would be larger in spherical shells with thinner convection zones. To test this idea we have run simulations with an aspect ratio of 0.6. In Fig.~\ref{Nu_ratio_TFSH_TNSH} we plot the $Nu$ ratio and free-slip $\varpi$ for this new data set. As expected the enhancement of convective efficiency in the no-slip case compared to the free-slip case is even higher in this setup, increasing about three fold in some cases. 

\section{Dynamo  simulations} \label{dynamo_sec}
In the previous section we showed that the presence of zonal flow is an important factor for the heat transfer efficiency in rotating convection and that the relative magnitude of the zonal jets depends on the mechanical boundary conditions. The magnetic field generated by a dynamo process can also, if it is strong enough, suppress zonal flows  \citep[see e.g.][]{aubert2005, miesch2005, browning2008, yadav2013a, heimpel2011, duarte2013, karak2015}. To explore this aspect we run self-consistent dynamo simulations for both kinds of mechanical boundary conditions.

\subsection{Dynamos with free-slip boundaries}

We begin by analyzing dynamo simulations with free-slip boundary conditions. Comparing these simulations with the corresponding HD simulations with free-slip boundaries will help us to pin down the change in heat-transfer efficiency due to the interaction of the magnetic field with the zonal-flows.

\subsubsection{Flow structure}

In Fig.~\ref{flow_struct}(e,f) we display the axial vorticity and the zonal flow profiles for dynamo simulations with free-slip boundaries. The zonal flow strength is reduced in Fig.~\ref{flow_struct}e as compared to its HD counterpart in Fig.~\ref{flow_struct}a. The reduction in zonal flow is less pronounced  for the high-$Ra$ case in Fig.~\ref{flow_struct}f. As a result of the reduced zonal shear, the vorticity structures seen in Fig.~\ref{flow_struct}e are more radially oriented in the equatorial plane. Both of these effects are qualitatively similar to what we already observed when we changed from free-slip to no-slip boundaries in HD simulations.

\begin{figure}
\includegraphics[scale=0.41]{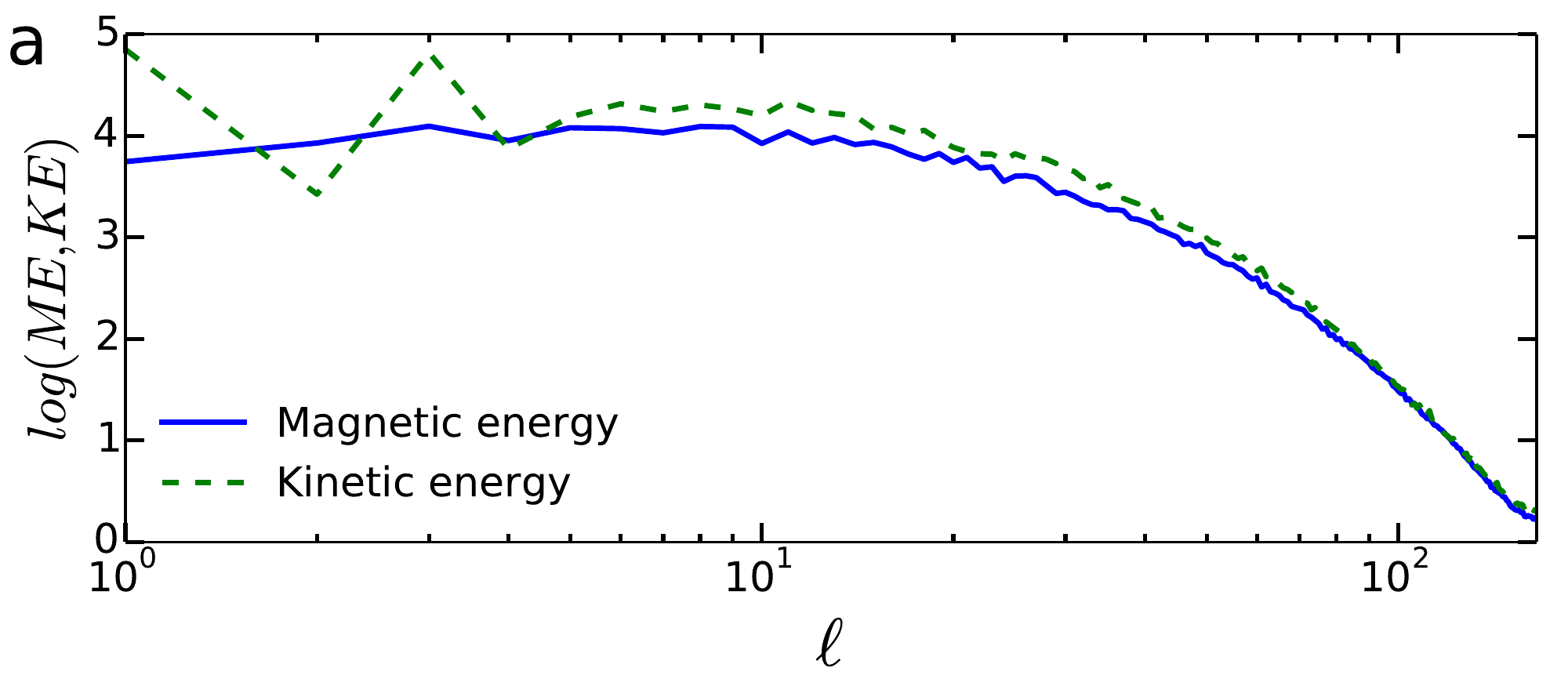} \includegraphics[scale=0.41]{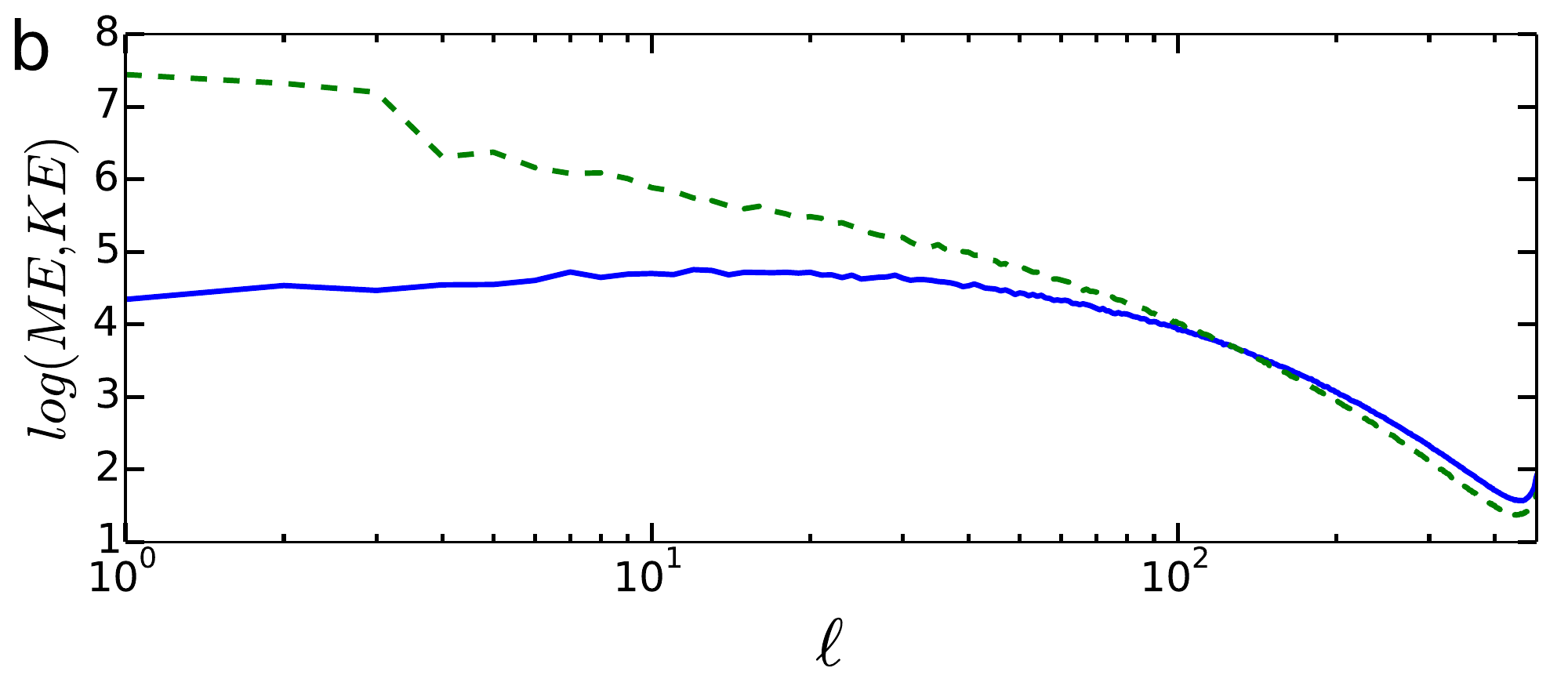} 
\caption{Spectrum of kinetic energy and magnetic energy as a function of the spherical harmonic degree $\ell$. The data is for a snapshot of the simulation. Panel a and b correspond to Fig.~\ref{flow_struct}e and \ref{flow_struct}f, respectively. \label{spectrum}}
\end{figure}

A feature similar to  the broad anticyclonic vorticity region seen in Fig.~\ref{flow_struct}b is visible again in Fig.~\ref{flow_struct}f. \cite{soderlund2013} have also reported such vorticity structures in longitudes (see their `thick-dynamo' case). The magnetic field is highly efficient at quenching the shear flow only when it is of sufficiently large scale and has energy comparable to or larger than the kinetic energy. In the dynamo simulations considered here the total magnetic energy (ME)  is close to the total kinetic energy for cases with $Ro_c<1$. This is due to the characteristic columnar convection which supports the generation of a large-scale dynamo. In this regime, the magnetic morphology could be either of geodynamo-type where the axial dipolar component carries the largest share of the total ME~\citep{kageyama1997, olson1999, aubert2008}, or it could be multipolar where higher-order field components dominate over the axial dipolar one~\citep{busse2006, christensen2006, brown2010, kapyla2012, gastine2012b}. Typically, in both kinds of dynamo solutions, the magnetic field at large scales remains comparable to the kinetic energy. Figure \ref{spectrum}a shows the comparable strength of the kinetic and the magnetic energy at different spherical harmonic degrees $\ell$ for the multipolar dynamo case shown in Fig.~\ref{flow_struct}e. Systematic studies of the  geodynamo have revealed that the magnetic energy becomes increasingly dominant as the Ekman number is decreased and can be an order of magnitude larger than the kinetic energy~\citep{christensen2006, takahashi2012}. Therefore, for lower $E$, the ratio ME/KE is likely to increase even further, enhancing the influence of the magnetic field on the flow. On the other hand, when $Ro_c$ is greater than unity, the helical columns are no longer dominant and plume-like convection generates a small-scale dynamo powered by the local chaotic motions. The spectrum shown in Fig.~\ref{spectrum}b, which corresponds to the case given in Fig.~\ref{flow_struct}f, highlights the much reduced magnetic energy at low $\ell$ or large length scales.  Therefore, we can expect that at small $Ro_c$, the back-reaction of the magnetic field on the flow will be rather strong, but as $Ro_c$ becomes larger, the dynamo becomes less efficient and the influence of the magnetic field would  become much weaker. This explains why the prograde zonal-jets are highly quenched for low $Ro_c$ dynamos while, for larger $Ro_c$, the retrograde zonal jets are not dramatically influenced~\citep{soderlund2013}.

\subsubsection{Latitude dependent Nusselt number}

The variation of heat flux with latitude for free-slip dynamo cases is shown in Fig.~\ref{lat_HF}b with dashed curves, along with the corresponding profiles for free-slip hydrodynamic cases. We again see that due to the quenching of the zonal flow by the magnetic field the heat transport at low latitudes is enhanced. However, unlike the HD simulations with no-slip boundaries, a significant enhancement of $Nu$ even inside the TC can be observed in Fig.~\ref{lat_HF}b.  The Nusselt number enhancement inside the TC only works for high enough $Ra$ since it is not present for lowest $Ra$ in Fig.~\ref{lat_HF}b. 

The dynamo simulation for which $Nu$ is enhanced inside the TC produces a dipole dominant magnetic field. This configuration has almost vertical poloidal magnetic field threading through the region inside the TC. We note that  dynamo simulations with free-slip boundaries exhibit an interesting properly called ``bistability" for a limited range of control parameters~\citep{simitev2009, schrinner2012, gastine2012b}. Such setups can sustain either a dipole dominant magnetic field or a multipolar magnetic field depending on the configuration and strength of the magnetic field used as initial condition. We exploit this property and generate a multipolar dynamo solution for the same control parameters as those used in generating the dipolar solution at $Ra=2\times10^8$. In Fig.~\ref{lat_HF}b we show the $Nu(\theta)$ profile for this multipolar dynamo solution using a trail of filled triangles. In this case $Nu$ is reduced inside the TC as oppose to an inhancement in the dipolar case shown with dashed curve. Close to the equator, the profiles are similar for dipolar and multipolar solutions. Therefore, for some simulations, the $Nu$ enhancement is due to two reasons: zonal flow quenching at low-latitudes and significant modifications of convection at high latitudes by the magnetic field.

\subsection{Relation between shear flow and heat transport}

\begin{figure}
\includegraphics[scale=0.48]{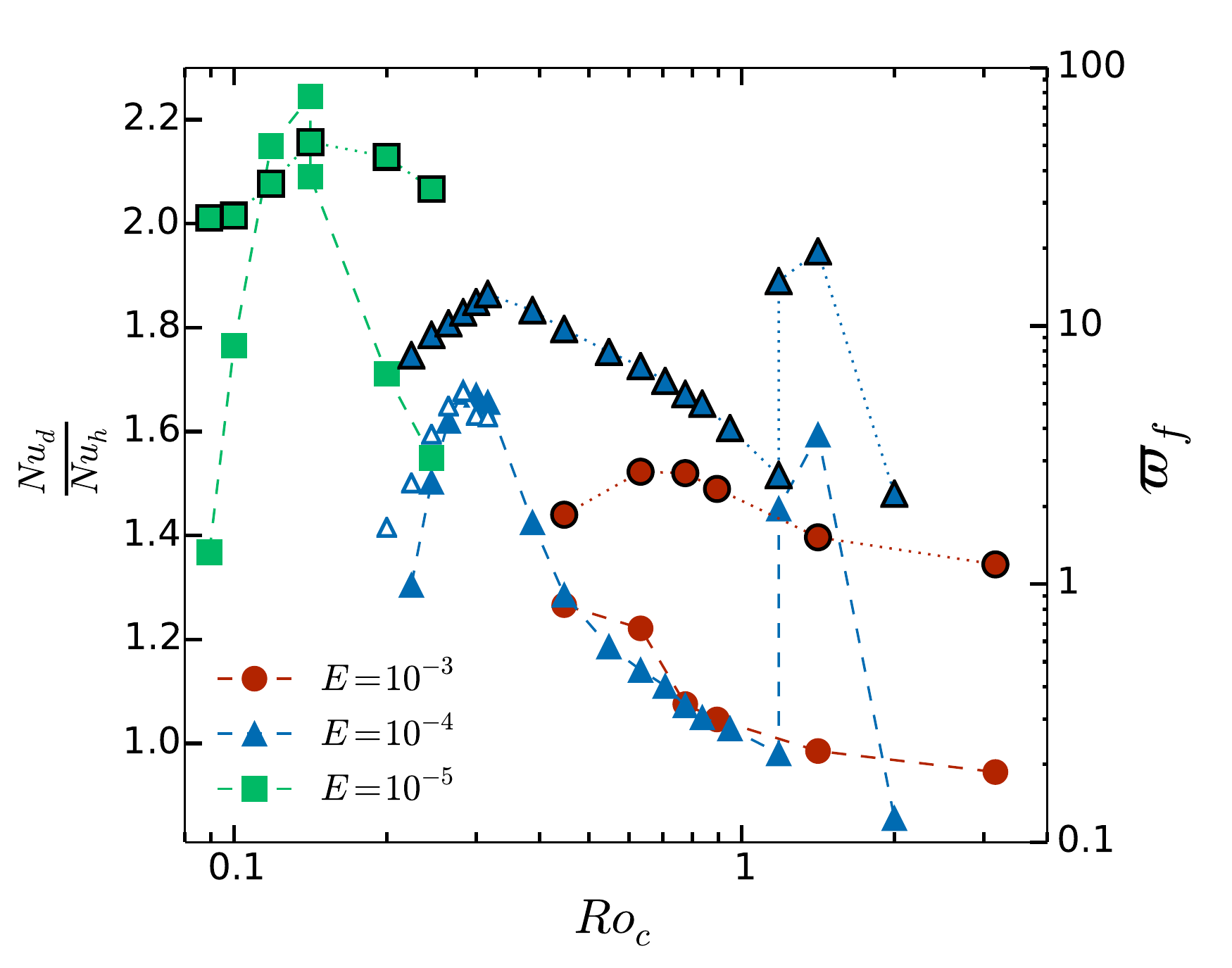}
\caption{On the left axis, ratio of the Nusselt number $Nu$ for free-slip dynamo (subscript `d') and free-slip HD (subscript `h') cases, and, on the right axis, $\varpi_{f}$ (black rimmed data points) for free-slip HD cases as a function of the convective Rossby number $Ro_c$.  The $Nu$-ratio for the bistable dipolar dynamos in the range $0.2<Ro_c<0.3$ is also plotted using small empty triangles. The data set is for simulations with aspect ratio of 0.35. \label{Nu_ratio_FSH_FSD}}
\end{figure}

Similar to Fig.~\ref{Nu_ratio_FSH_NSH} we compare the $Nu$ ratio for free-slip HD and free-slip dynamo simulations and $\varpi$ for free-slip HD simulations in Fig.~\ref{Nu_ratio_FSH_FSD}. Once again a good correlation exists between the two quantities, re-enforcing the idea that quenching of zonal jets is the primary cause of the heat-transfer efficiency enhancement in the dynamo cases compared to the non-magnetic cases. 

For a Rossby number larger than unity the Nusselt number of dynamo simulations is somewhat smaller than that of the corresponding HD simulations. This indicates that the small-scale dynamo operating in the high $Ro_c$ regime is introducing the magneto-quenching effect where the presence of the magnetic field makes the convection less efficient at transferring heat. Since the magnetic field itself is weak in this regime, the associated decrease in $Nu$ is also rather small.

It is possible that the magnetic quenching effect is present even for $Ro_{c}<1$. In this case, even though the magnetic field may  quench convection to some extent, the increase in $Nu$ due to the reduced zonal flow is much larger than the reduction in $Nu$ due to magnetic quenching. For larger $Ro_{c}$, on the other hand, since the effect of zonal flow is much weaker, the reduction in $Nu$ by magnetic quenching prevails.

\subsection{Dynamos in deep convective layers}

We also compare the heat-transfer efficiency in HD and dynamo simulations in spherical shells with very deep convection zones (shells with aspect ratio of 0.2). For this purpose we analyse dynamo cases from \citet{gastine2012b}, supplemented by additional similar hydrodynamic simulations. These cases are not completely analogous to our basic setup: the inner boundary is no-slip while the outer boundary is free-slip. Since the non-convective inner core is rather small its boundary condition is unlikely to strongly affect the dynamics of the global convective layer. 

\begin{figure}
\includegraphics[scale=0.48]{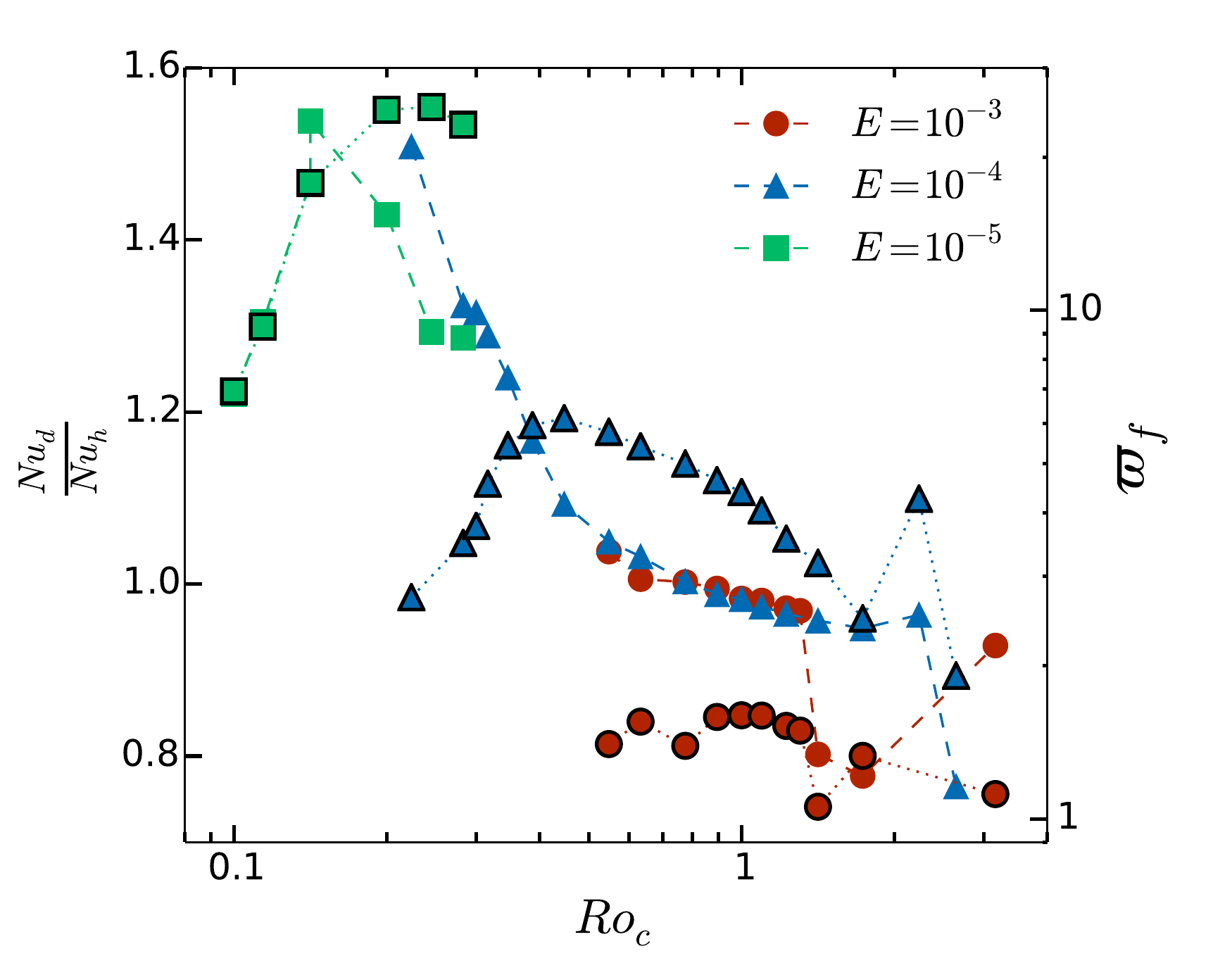}
\caption{On the left axis, ratio of the Nusselt number $Nu$ for free-slip dynamo (subscript `d') and free-slip HD (subscript `h') cases, and, on the right axis, $\varpi_{f}$ (black rimmed data points) for free-slip HD cases as a function of the convective Rossby number $Ro_c$. The data set is for simulations with aspect ratio of 0.2.   \label{Nu_ratio_LFSD_LFSH}}
\end{figure}

We plot $Nu$ ratio and $\varpi$ in Fig.~\ref{Nu_ratio_LFSD_LFSH}. The contribution of zonal flow is reduced in this thick-shell configuration as compared to the earlier thinner spherical shells. For instance, the parameter $\varpi_{h}$, defining the relative stregnth of the zonal flow, reaches a maximum value of about 25 at $E$=$10^{-5}$ in the HD simulations with $\eta$=0.2 while it reached about 50 in the HD cases with $\eta$=0.35 and $E$=$10^{-5}$ (Fig.~\ref{Nu_ratio_FSH_FSD}). It shows that the zonal flow strength is weaker in the thick shell cases. For $E$=$10^{-3}$ cases, there is no correlation between the $Nu$ ratio and $\varpi_{h}$. For $E$=$10^{-4}$ a correlation exists but only for large enough $Ro_c$, and, at $E$=$10^{-5}$, a tight correlation for low $Ro_c$ and a rough correlation at higher $Ro_c$ appears. Therefore, as we decrease the Ekman number, the energy content of the zonal flow increases and we start to see a better correlation between  $Nu$ ratio and $\varpi_{h}$. Nonetheless, comparing this data-set with that for $\eta=0.35$ (Fig.~\ref{Nu_ratio_FSH_FSD}) indicates that spherical shells with deep convective layers promote weaker zonal flows as compared to thinner spherical shells. The weaker enhancement in $Nu$ due to the presence of a dynamo-generated magnetic field follows as a consequence.

\subsection{\label{sec:NSD_subsec}Dynamos with  no-slip boundaries}
In the earlier sections we saw that either the no-slip mechanical boundary conditions or the presence of a dynamo-generated magnetic field enhance the convective efficiency by quenching the zonal shear. The magnetic field, however, is special in that it not only affects the zonal shear but also changes the structure of the convective flow responsible for transferring heat. In this sub-section we investigate dynamo simulations which have no-slip flow boundaries. Comparing the no-slip HD simulations with these dynamo simulations will allow us to minimize the effects of different zonal flow amplitudes on the Nusselt number.

\subsubsection{Flow structure}
The axial vorticity plots and the zonal flow profiles are shown in Fig.~\ref{flow_struct}(g,h) for these new dynamo cases. The zonal flows are highly quenched, and, correspondingly, the shearing in the axial vorticity structures is much weaker. The zonal flow also varies substantially along the rotation axis, usually observed in dynamos with dipole-dominant magnetic field configurations~\citep{aubert2005, schrinner2012, yadav2013a}. In the high $Ra$ regime (small-scale dynamo cases), the zonal flow profile is similar to the corresponding hydrodynamic case in Fig.~\ref{flow_struct}d. Once again it illustrates that small-scale dynamos have a much smaller influence on the flow as compared to the dipolar or the multipolar dynamos at smaller $Ro_c$ (at least in our simulations).

\subsubsection{Latitude dependent Nusselt number}
The latitudinal heat-flow profiles for no-slip dynamo cases are shown by dashed curves in Fig.~\ref{lat_HF}c, along with the corresponding profiles for no-slip HD cases. Near the equator both HD and dynamo cases behave somewhat similarly. This is due to the fact that the zonal jets, responsible for controlling the heat-transfer at low-latitudes in the free-slip cases, are highly quenched in both setups. Inside the TC, however, different behaviour exist at different Rayleigh numbers: similar heat flow for $Ra$=$8\times10^7$ (red curves), increased heat flow in the dynamo case at $Ra$=$2\times10^8$ (blue curves), and mixed behavior for $Ra$=$6\times10^8$ (green curves). The sharp increase close to the poles in Fig.~\ref{lat_HF}c appears to be due to the presence of large-scale helical polar vortices with enhanced radial and horizontal flows as compared to the corresponding HD case. Such vortices are preferably excited when a strong dipolar field is present~\citep{sreenivasan2005}.

\subsubsection{Relation between field strength and heat transport}
\begin{figure}
\includegraphics[scale=0.46]{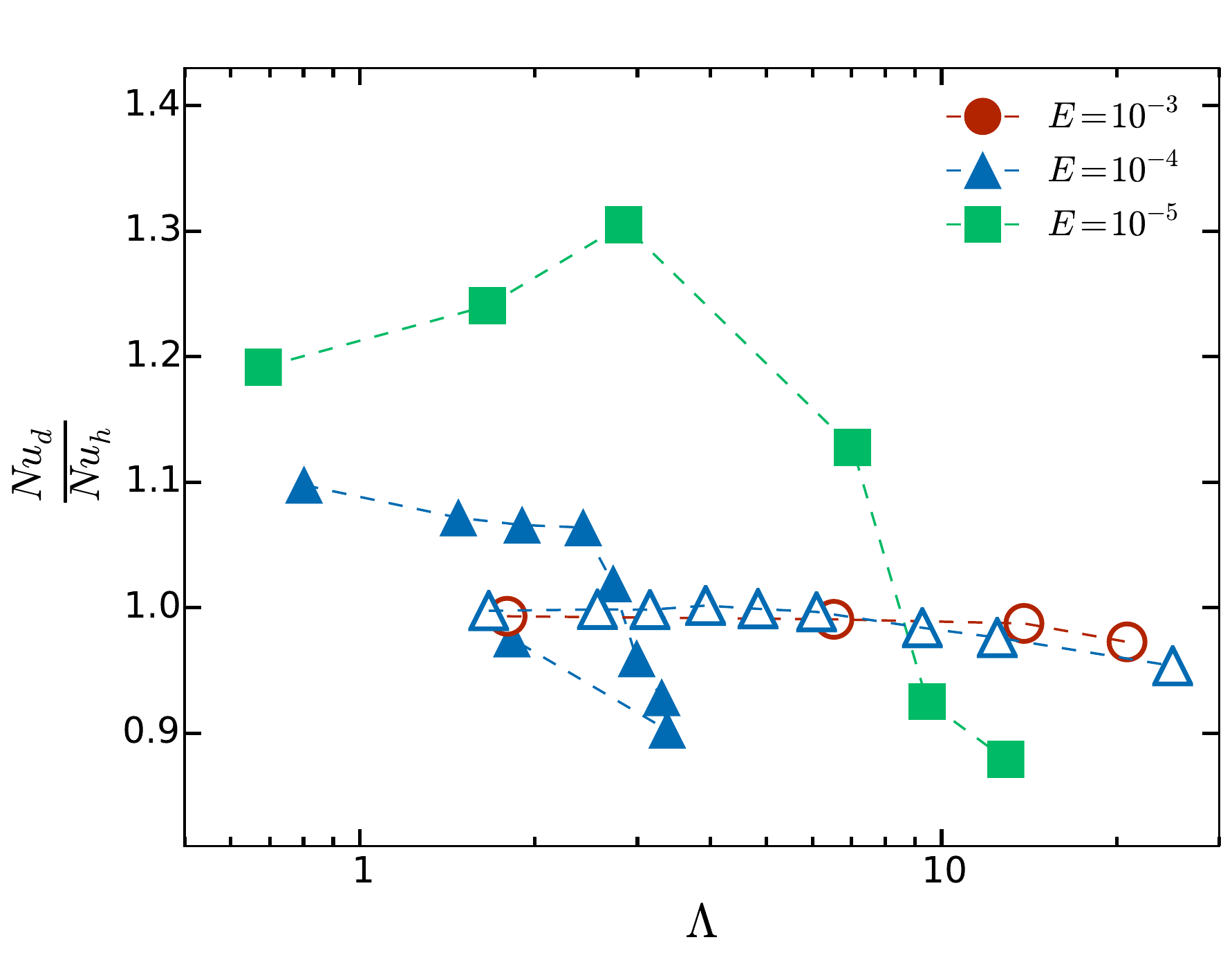}
\caption{Variation of the ratio of $Nu$ for dynamo (subscript `d') and hydrodynamic (subscript `h') cases with no-slip boundaries as a function of the Elsasser number $\Lambda$. Filled/unfilled symbols represent dipolar/multipolar dynamos. Only dynamo simulations with $P_m$=1 are shown. \label{Nu_ratio_Els}}
\end{figure}

In both HD as well as dynamo simulations with no-slip boundaries the energy in the zonal flow is rather small. Therefore, the zonal flows do not play an important role in determining the heat-transfer efficiency in these setups. In Fig.~\ref{Nu_ratio_Els} we compare the ratio of $Nu$ for dynamo and HD cases as a function of the mean magnetic field strength ($B$) which is quantified by the classical non-dimensional parameter called the Elsasser number 
\begin{gather}
\Lambda = \frac{B^2}{\rho\mu\lambda\Omega}.
\end{gather}
This figure shows two distinct features. For the multipolar dynamos (empty symbols), the magnetic field only marginally quenches the heat transport. For the dipolar dynamos at $E$=$10^{-5}$, however, a peak is present at $\Lambda\approx 3$ where the efficiency of convection in transporting heat is enhanced by 30\%. At a higher Ekman number of $10^{-4}$ the enhancement of the Nusselt number is smaller and reverses for $\Lambda > 3$. Note that the dashed line for $E$=$10^{-4}$ data points follows the increasing $Ro_{c}$ trend. When the dynamo changes behavior from dipolar to multipolar the field strength decreases eventhough $Ro_{c}$ is increased. Therefore, the blue dashed-line does not follow a strictly increasing trend.

The shape of the green curve (for $E$=$10^{-5}$) is quite similar to that reported by \cite{king2015}. In this study liquid Gallium ($P_r$ $\approx$ 0.025 and $P_m$ $\approx$ $10^{-6}$) was used as the working fluid and rotating convection ($2\times10^{-6}\le E \le  2\times10^{-4}$) was investigated in a cylindrical container with/without a vertically imposed magnetic field. This study reported maximal convective efficiency (about 50\% higher $Nu$ in magnetoconvection) at $\Lambda\approx 2$ and $E\approx10^{-6}$. The authors interpret the $Nu$-enhancement as a signature of `magnetostrophic' regime of convection~\citep{chandrasekhar1954, stevenson1979}. Rotating convection in cylindrical/Cartesian setups with vertically imposed magnetic field can be compared with the polar-region convection in rotating spherical-shells generating a dipole-dominant magnetic field.

The similarity of $Nu$ enhancement in our $E$=$10^{-5}$ cases with the experimental results by \citet{king2015} suggests that magnetic field is playing a dominant role in governing the convection behavior in the no-slip dynamo cases, especially at high latitudes (Fig.~\ref{lat_HF}c). In this context, it is interesting to note a recent study by \citet{teed2015} who analysed torsional oscillations in a magnetoconvection setup in rotating spherical shells and find a possible transition to magnetostrophic regime also at $E\approx 10^{-5}$. This hints that such Ekman number simulations might be entering the magnetostrophic convection regime. However, without a detailed analysis of various forces in our simulations we can not confirm the existence of such a regime. We defer this exercise to our forthcoming study where lower Ekman number simulations will be analysed.

\section{Summary and outlook} \label{sum}

In this study we sought to shed some light on the question of what factors affect the heat-transfer efficiency of convection in rotating spherical shells  as it occurs in fluid regions of planets and stars. We compare four setups, hydrodynamic simulations with either free-slip or no-slip flow boundary conditions and the corresponding self-consistent dynamo cases. The key features can be summarised as follows:

\begin{itemize}
\item Hydrodynamic simulations with free-slip (stress-free) boundaries generate the strongest zonal jets. These axially-aligned zonal flows disrupt and impede the heat-transporting motions. Their effect is most efficient in the equatorial regions where they severely quench the efficiency of convection, whereas, in the polar regions, the effect of the zonal flows on the heat-transport efficiency was substantially reduced. 

\item Changing the free-slip boundary condition to no-slip strongly reduces the energy of the zonal flows. Consequently, the Nusselt number increases substantially, by as much as a factor of 3 in some of our simulations with a thin convecting shell. The heat transport is most strongly enhanced at low latitudes. However, in the inertia or buoyancy dominated regime at large Rossby numbers, the Nusselt number is slightly higher with free-slip boundaries.

\item The presence of a dynamo-generated magnetic field in the case of free-slip boundaries also efficiently quenches the zonal flows and enhances the heat flow at low latitudes. The heat-flux in the polar regions is affected by the magnetic field as well. Dipolar dynamos can substantially increase the polar heat-flux, whereas multipolar dynamos can quench it. Again, this holds in the rotation-dominated regime at low Rossby numbers.

\item In case of no-slip boundaries, when zonal flows play a minor role also without a magnetic field, the presence of a dynamo-generated field can enhance the Nusselt number significantly. Conditions for this to happen are that the Ekman number is low enough ($E\leq 10^{-5}$) and the magnetic field is dipole-dominated and has a suitable strength, that is, the Elsasser number $\Lambda$ is of order one. At $E=10^{-5}$ the optimum is reached for $\Lambda\approx 3$.

\item Our results demonstrate the importance of self-excited zonal flows and magnetic fields for the heat transfer efficiency of rotating convection, as well as the different influence that they have inside and outside the tangent cylinder. This suggests that studies of heat transfer in Cartesian layers~\citep[for instance,][]{barker2014} or cylindrical geometry~\citep[for instance,][]{cheng2015} may miss many crucial effects that are important in planetary or stellar applications.

\end{itemize}

There are many fronts where our study can be extended. We have not assessed the influence of the (magnetic) Prandtl number on our results. Increasing and decreasing the fluid Prandtl number has a direct effect on the strength of the zonal flows in rotating shells, with zonal flows being stronger for $P_r < 1$~\citep[e.g.][]{zhang1992, aubert2001, christensen2002,  simitev2005, gillet2007, calkins2012}. The magnetic Prandtl number might also be important since it affects the field strength and morphology of the dynamo-generated magnetic field~\citep[e.g.][]{christensen2006, schrinner2012, yadav2013a, yadav2013b, raynaud2014}. Compressibility, which is very important for convection in giant planets and stars, is  another aspect which likely influences the convective efficiency by changing the nature of the zonal flows~\citep{gastine2012a}. Exploration of these aspects and testing the approach towards a magnetostrophic flow regime by simulations at even lower values of the Ekman number are essential steps towards a better understanding of rotating and magnetic convection in natural objects.

\section*{acknowledgements}
We thank the reviewers, Julien Aubert and Robert Teed, for many useful comments which greatly improved the presentation of the manuscript. We also thank Jonathan Aurnou for interesting comments. We acknowledge support from  the  Deutsche Forschungsgemeinschaft (DFG) through Project SFB 963/A17 and the Special Priority Program 1488. RKY also acknowledges support from NASA {\em Chandra} grant GO4-15011X. Computations were performed on the GWDG computer facilities in G\"ottingen, Regionales Rechenzentrum f\"ur Niedersachsen, HLRN under project ``nip00031", and at Rechenzentrum Garching (RZG). We also thank Tilman Dannert (RZG) for implementing MPI and then hybrid OpenMP+MPI parallelisation in MagIC.

\bibliographystyle{natbib}

\bibliography{cited}

\label{lastpage}

\end{document}